\documentclass[aps,pra,reprint,showkeys,amssymb,floatfix,longbibliography,superscriptaddress]{revtex4-2}

\usepackage{graphicx}
\usepackage{amsmath,amsfonts,amssymb}
\usepackage{hyperref}
\usepackage{braket}
\usepackage{xcolor}
\usepackage{tikz}
\usepackage{color}
\usepackage{enumerate}
\usepackage[caption=false]{subfig}
\usepackage{multirow}
\usepackage{qcircuit}
\usepackage[normalem]{ulem}
\usepackage{placeins}
\usepackage{xurl}
\usepackage{dsfont}
\usepackage{braket}
\usepackage[multidot]{grffile}
\usepackage{mathtools}
\usepackage{float}
\usepackage{textcomp}

\newcommand{\equref}[1]{Eq.~(\ref{#1})}
\newcommand{\figref}[1]{Fig.~\ref{#1}}
\newcommand{\secref}[1]{Sec.~\ref{#1}}

\begin{document}

\title{Guided quantum walk}

\author{Sebastian Schulz}
\thanks{Corresponding author: Sebastian Schulz}
\email{se.schulz@fz-juelich.de}
\affiliation{J\"ulich Supercomputing Centre,
Institute for Advanced Simulation,\\
Forschungszentrum J\"ulich, 52425 J\"ulich, Germany}
\author{Dennis Willsch}
\affiliation{J\"ulich Supercomputing Centre,
Institute for Advanced Simulation,\\
Forschungszentrum J\"ulich, 52425 J\"ulich, Germany}
\author{Kristel Michielsen}
\affiliation{J\"ulich Supercomputing Centre,
Institute for Advanced Simulation,\\
Forschungszentrum J\"ulich, 52425 J\"ulich, Germany}
\affiliation{AIDAS, 52425 J\"ulich, Germany}
\affiliation{RWTH Aachen University, 52056 Aachen, Germany}

\date{\today}

\begin{abstract}
We utilize the theory of local amplitude transfer (LAT) to gain insights into quantum walks (QWs) and quantum annealing (QA) beyond the adiabatic theorem. By representing the eigenspace of the problem Hamiltonian as a hypercube graph, we demonstrate that probability amplitude traverses the search space through a series of local Rabi oscillations. We argue that the amplitude movement can be systematically guided towards the ground state using a time-dependent hopping rate based solely on the problem's energy spectrum. Building upon these insights, we extend the concept of multi-stage QW by introducing the guided quantum walk (GQW) as a bridge between QW-like and QA-like procedures. We assess the performance of the GQW on exact cover, traveling salesperson and garden optimization problems with 9 to 30 qubits. Our results provide evidence for the existence of optimal annealing schedules, beyond the requirement of adiabatic time evolutions. These schedules might be capable of solving large-scale combinatorial optimization problems within evolution times that scale linearly in the problem size.
\end{abstract}


\maketitle


\section{Introduction}

Combinatorial optimization is a fundamental problem in computer science that has a wide range of important applications in finance \cite{PortfolioOpt_Marzec}, scheduling \cite{SC_Willsch}, machine learning \cite{ML_Amin, ML_Benedetti}, database search \cite{Childs2004CTQWSearchProblem}, computational biology \cite{CB_Perdomo}, and operations research~\cite{OperationsResearch}. However, finding optimal solutions to such problems can be challenging and computationally expensive, which often makes them intractable for classical computers today. Recent advances in quantum hardware are raising the expectations for demonstrating useful quantum computation in the upcoming years. In particular, solving large-scale combinatorial optimization problems is considered one of the great application areas of quantum computation, driving the need for quantum optimization algorithms suitable for near-term quantum devices~\cite{Preskill2018NISQ}.

Quantum walks (QW)~\cite{Aharonov1993QuantumRandomWalks,Childs2002QuantumSearchByMeasurement,Childs2004CTQWSearchProblem,Callison2021EnergeticPerspectiveQAMultiStageQW,banks2023CTQWforMAXCUTarehot} and quantum annealing (QA)~\cite{Apolloni89,finnila94,KadowakiNishimori1998QuantumAnnealing,Brooke99,Johnson2011DWave,Harris2010DWave,Bunyk2014DWave,AlbashLidar2018AdiabaticQuantumComputation,Albash2018QuantumAnnealingScalingAdvantageOverSimulatedAnnealing,King2023QuantumCriticalDynamics5000SpinGlass} have emerged as two promising candidates for continuous-time quantum optimization algorithms in this context. QWs, introduced by Aharonov et al.~\cite{Aharonov1993QuantumRandomWalks}, model the search space as a graph and govern the walker's transitions between vertices using a time-independent Hamiltonian. On the other hand, QA employs a time-dependent Hamiltonian to adiabatically evolve from an initial Hamiltonian to a problem Hamiltonian.~Both algorithms have been extensively studied for various problems with tens of qubits, including Sherrington-Kirkpatrick spin glasses~\cite{FindingSpinglas_Callison, Callison2021EnergeticPerspectiveQAMultiStageQW}, Max-Cut \cite{banks2023CTQWforMAXCUTarehot}, 2-SAT~\cite{neuhaus2014monteCarloQA2SAT,neuhaus2014quantumSearches2SAT,Mehta2021quantumAnnealingWithTriggerHamiltonians2SATNonstoquastic,mehta2022QuantumAnnealingForHard2SATProblems,mehta2022hardnessOfQUBO} and exact cover \cite{SC_Willsch} problem instances. While these algorithms exhibit distinct dynamics, Morley et al.~\cite{HybridQAQW_Morley} have argued that they can be seen as extreme cases of annealing schedules in the context of search problems. However, the dynamics occurring in the intermediate region between QA and QW for combinatorial optimization problems have yet to be fully explored. We argue that these intermediate evolutions, which go beyond the scope of the adiabatic theorem, are a very promising regime for effective quantum computation.

In this paper, we utilize the theory of local amplitude transfer (LAT) to investigate continuous-time quantum algorithms. LAT theory focuses on the local energy structure of the problem Hamiltonian on the hypercube graph. It provides insights into the movement of probability amplitude between individual elements of the search space through a series of local Rabi oscillations. 

Building upon these insights, we extend the concept of multi-stage QW~\cite{Callison2021EnergeticPerspectiveQAMultiStageQW,banks2023CTQWforMAXCUTarehot} by introducing the guided quantum walk (GQW). The GQW combines multiple QWs through a time-dependent hopping rate, effectively guiding the transfer of probability amplitude throughout the graph. This approach relates to the problem of finding optimal annealing schedules~\cite{QAOpt_PathOptZeng, QAOpt_PathOptHerr, QAOpt_MonteCarloTreeSearch}, but goes beyond the requirement of adiabatic time evolutions, placing the GQW in between QW-like and QA-like procedures.

To evaluate the performance of the GQW, we numerically simulate its application to exact cover (EC)~\cite{Karp1972KarpsNPCompleteProblems,EC_Choi,SC_Willsch}, traveling salesperson (TSP)~\cite{Dantzig1954SolutionOfALargeScaleTSP,Bellman1962TSPDynamicProgramming,HeldKarp1962TSP,OperationsResearch,Lucas2014IsingQUBOFormulationManyNPproblems}, and garden optimization (GO)~\cite{calaza2021gardenoptimization} problems using the JUWELS Booster supercomputer~\cite{JuwelsClusterBooster}. We extensively study the GQW on problem instances with up to 30 qubits. Our results provide evidence for the existence of optimal annealing schedules capable of solving large-scale combinatorial optimization problems with evolution times that scale linearly in the problem size.

\clearpage

The paper is organized as follows: In \secref{sec:comb_problems}, we introduce the types of combinatorial optimization problems that are used in our research. Section~\ref{sec:quantum_walk} provides a review of QWs and explores their dynamics on search and optimization problems using the LAT theory. We discuss how probability amplitude can be effectively guided through the hypercube graph, leading to the development of the GQW. Furthermore, we examine the relationship between the GQW, QA and QWs at different evolution times. In \secref{sec:results}, we present a comprehensive performance analysis of the GQW. Finally, in \secref{sec:conclusion}, we summarize our findings and their implications for future research.


\section{Combinatorial Optimization Problems}
\label{sec:comb_problems}
This section presents the combinatorial optimization problems studied in our research. First, we describe the problem types and their encoding into a quantum setting in \secref{subsec:comb_problems:definition}. Subsequently, we introduce a benchmarking metric, the solution quality, which ensures fair comparisons of quantum optimization algorithms across different problem types and sizes in \secref{subsec:comb_problems:Sq}.

\subsection{Definition}
\label{subsec:comb_problems:definition}
A broad class of real-world problems can be framed as combinatorial optimization problems~\cite{OperationsResearch}. These are problems defined on N-bit binary strings $z = z_{N-1} \dots z_{0} = \left\{0, 1\right\}^{\otimes N}$, with the objective of finding a string $z_{opt}$ that minimizes a given classical cost function $C\left(z\right) : \left\{0, 1\right\}^{\otimes N} \xrightarrow{} \mathbb{R}_{\geq 0}$. A natural way to express $C$ in a quantum setting is to encode it into the energy spectrum of the computational basis states $\ket{z}$ of a quantum cost Hamiltonian
    \begin{align}
        \hat{H}_C = \sum_{z \, \in \, \left\{0, 1\right\}^{\otimes N}} C\left(z\right) \ket{z}\bra{z}.
        \label{eq:H_C}
	\end{align}
We consider the case in which $C\left(z\right)$ consists only of linear and quadratic terms, $C\left(z\right)=\sum_{ij} z_i Q_{ij} z_j$. Such problems are known as quadratic unconstrained binary optimization (QUBO) problems, with $\{Q_{ij}\}$ denoting the QUBO coefficients. For these problems, $\hat{H}_C$ can be written in the form of an Ising Hamiltonian by substituting $z_i\mapsto (1-\hat{\sigma}_i^z)/2$, where $\hat{\sigma}_i^z$ denotes the Pauli-\textit{Z} operator applied to the \textit{i}th qubit with the identity operator acting on the remaining qubits. One obtains
    \begin{align}
        \hat{H}_C = -\sum_{i=0}^{N-1} h_i \hat{\sigma}_i^z + \sum_{i\neq j} J_{ij}\hat{\sigma}_i^z\hat{\sigma}_j^z,
        \label{eq:H_CIsing}
	\end{align}
with the coefficients $h_i=\sum_j(Q_{ij}+Q_{ji})/2$ and $J_{ij}=Q_{ij}/4$ representing the optimization problem. In our study, we investigate three types of combinatorial optimization problems, which can be expressed in Ising form and represent two categories of cost functions.

The first category (constraint-only) concerns cost functions consisting entirely of constraining terms, meaning that all states describing valid solutions to the optimization problem (called valid states henceforth) are assigned to the same cost value. Here, we focus on EC problem instances with $N \in \left\{12, 15, 18, 21, 24, 30\right\}$. Generically, these problem instances exhibit numerous distinct energy levels with low degeneracies and a non-degenerate ground state.

The second category (constraint+optimization) includes cost functions that involve both constraining and optimization terms, such that the set of valid states spans multiple energy levels. We consider TSP and GO problem instances with $N \in \left\{9, 16, 25\right\}$ and $N \in \left\{12, 15, 18, 21, 24, 30\right\}$, respectively. It is worth noting that the GO instances generally exhibit higher degeneracies among their energy levels compared to the TSP instances.

The explicit cost functions for the three types of combinatorial optimization problems can be found in App.~\ref{app:cost_functions}. For each problem type and size, we investigate $10$ randomly generated problem instances. Note that all cost functions have been rescaled and shifted, such that $C\left(z_{opt}\right) = 0$ and $\max_{z} C\left(z\right) = 100$.

\subsection{Solution quality}
\label{subsec:comb_problems:Sq}
Quantum optimization algorithms are designed to find the optimal solution to a given combinatorial optimization problem (\equref{eq:H_C}) with high success probability $P_{z_{opt}} = \left|\braket{z_{opt}|\Psi}\right|^2$, where $\ket{\Psi} = \sum_z \psi_z \ket{z}$ denotes the final quantum state. However, in many cases, approximate solutions, where not solely the solution state but also other valid states with a slightly larger cost function value are obtained with high probability, are also of interest, especially if they can be found significantly faster. Since $P_{z_{opt}}$ does not take these approximate solutions into account, a common approach is to evaluate the performance of quantum optimization algorithms based on the energy expectation value
    \begin{align}
        E_{\Psi} = \braket{\Psi | \hat{H}_C | \Psi},
        \label{eq:<H_C>}
	\end{align}
often in the form of the approximation ratio $r\left(\Psi\right) = E_{\Psi} / E_{max}$, where $E_{max}=\max_{z} E_{z}$. Note that smaller approximation ratios are supposed to represent better solutions. 

We emphasize, however, that the approximation ratio $r$ in \equref{eq:<H_C>} is often not able to capture the 'quality' of the produced quantum state $\ket{\Psi}$, that is the distribution of the measurement probabilities w.r.t.~the energies of the valid states. This is because $r$ considers not only the set of valid states (which we are generally interested in) but also the set of invalid states (i.e.~states that violate at least one constraint and may shift $E_{max}$ arbitrarily). For the problems under investigation, the latter correspond to the majority of states, covering $\geq 95\%$ of the total energy range. Hence, comparing the approximation ratio between different quantum optimization algorithms does not necessarily compare their ability to produce 'good' approximate solutions, as a final state with a large value of $r$ might still provide valid states near the optimal solution with higher probability than a different state of smaller $r$ value. 

In order to address this issue of the approximation ratio, we propose the use of a theoretical benchmarking metric termed the \textit{solution quality}
    \begin{align}
        S_q = \sum_{z \, \in \, \text{valid states}} \left[1 - r_{valid}\left(z\right) \right] \cdot P_z,
        \label{eq:S}
	\end{align}
where $P_z=\lvert\braket{\Psi | z}\rvert^2$ represents the measurement probability of the state $\ket{z}$, and $r_{valid}\left(z\right) = E_z / \max_{k \, \in \, \text{valid states}}(E_k)$ is used to give a weight among all valid states in the energy spectrum (note that $\ket{z_{opt}}$ is at $E_{z_{opt}} = 0$ by definition). $S_q$ is designed to capture both the characteristics of the success probability and the approximation ratio, restricted to the set of valid states. This provides a theoretical benchmarking metric that is comparable across different problem instances by focusing on the practicality of the solutions obtained. Note that $S_q \in \left[0, 1\right]$, with $S_q = 1$ corresponding to the solution state $\ket{\Psi} = \ket{z_{opt}}$, and $S_q = 0$ indicates a final state $\ket{\Psi}$ that consists solely of invalid states and highest-energy valid states with $r_{valid} = 1$. The latter are effectively excluded from increasing the solution quality $S_q$, because obtaining any valid solution to the constraint+optimization problems (TSP and GO) can be done in polynomial time. Hence, we do not consider an information gain from these states. Furthermore, in the case of constraint-only problem instances (EC), we set $r_{valid}\left(z\right) = 0$, such that $S_q = P_{z_{opt}}$.


\section{Continuous-time quantum walk}
\label{sec:quantum_walk}
In this section, we provide a concise review of the continuous-time QW (\secref{subsec:qw:definition}) and investigate its dynamics on search (\secref{subsec:qw:search_problem}) and combinatorial optimization problems (\secref{subsec:qw:optimization_problem}) using LAT theory. Our analysis shows that the QW can be systematically guided on the hypercube graph based on the energy spectrum of the problem Hamiltonian. Building upon these insights, we introduce the GQW in \secref{subsec:qw:guided_quantum_walk}. In \secref{subsec:qw:relation_qa} we examine the relationship between the GQW, QA and QWs.

\subsection{Definition}
\label{subsec:qw:definition}
The continuous-time QW is a quantum algorithm that assigns the computational basis states $\ket{j}$ of an $N$-qubit Hilbert space to the set of vertex labels $V=\left\{j\right\}_{j = 0}^{2^N-1}$ of an undirected graph $G\left(V, T\right)$. In this framework, the vertices encode the walker's position, and the set of edges $T$ indicate allowed transitions between label pairs $\left(j, k\right)$. The latter is described through an adjacency matrix $A$, whose elements satisfy $A_{j, k} = 1$ if an edge in $G$ connects vertices $j$ and $k$, and $A_{j, k} = 0$ otherwise. As $G$ is undirected, $A$ is symmetric and can be used to define the quantum-walk Hamiltonian, given by:
    \begin{align}
        \hat{H}_{QW} = \Gamma \cdot \hat{H}_D = -\Gamma \cdot \sum_{j, k} A_{j, k} \ket{j}\bra{k},
        \label{eq:H_QW}
	\end{align}
where $\hat{H}_D$ denotes the driver Hamiltonian, and $\Gamma$ is the hopping rate. It is important to note that \equref{eq:H_QW} is not the only possible Hamiltonian for a QW. In the literature, the Laplacian $\hat{L} = -\hat{H}_D - \hat{D}$ of $G$ is often used instead of $\hat{H}_D$, with the diagonal matrix $\hat{D} = \sum_j \text{deg}(j) \ket{j}\bra{j}$ encoding the degree $\text{deg}(j)$ of each vertex $j$, i.e.~the number of edges incident to $j$~\cite{Childs2004CTQWSearchProblem}. However, for the regular graphs $G$ considered in this paper, where $\text{deg}(j)$ is constant with respect to $j$, both formulations are equivalent up to an unobservable global phase factor. We chose to use the adjacency operator form of $\hat{H}_{QW}$ for consistency with other quantum optimization algorithms.

Given the QW Hamiltonian (\equref{eq:H_QW}), the state of the walker evolves from some initial state $\ket{\Psi_0}$ according to the time-dependent Schrödinger equation, which yields the state of the system at a time $T$ as
    \begin{align}
        \ket{\Psi\left(T\right)} = e^{-i \hat{H}_{QW} T} \ket{\Psi_0},
        \label{eq:QW_TDSE}
	\end{align}
where we have used units with $\hbar = 1$. The quantum dynamics implemented by this evolution clearly depend on the connectivity in the graph $G$. In the past, QWs have been studied on a variety of graph layouts, including the complete graph, which couples every vertex to every other, and the $N$-dimensional hypercube, which connects only vertices of Hamming distance one \cite{Childs2004CTQWSearchProblem}. In this paper, we focus on a hypercube, as it provides a natural encoding for a QW into qubits. Specifically, transitioning from one vertex to a neighboring vertex corresponds to flipping the computational state of one qubit. As such, the driving Hamiltonian $\hat{H}_D$ is composed of $N$ single-body terms,
    \begin{align}
        \hat{H}_D = -\sum_{j=1}^{N} \hat{\sigma}_j^x,
        \label{eq:H_D_Hypercube}
	\end{align}
where $\hat{\sigma}_j^x$ denotes the Pauli-X operator applied to the \textit{j}th qubit with the identity operator acting on the remaining qubits. The corresponding QW Hamiltonian for the hypercube is given by $\hat{H}_{QW} = -\Gamma \cdot \sum_{j=1}^{N} \hat{\sigma}_j^x$. A primary feature of this Hamiltonian is its ability to rapidly explore the vertices in the graph $G$, thereby providing high dynamics in the computational basis. By introducing a secondary problem Hamiltonian that is diagonal in the computational basis, the graph $G$ becomes directed, leading to a concentration of amplitude in the ground state of the problem Hamiltonian (see App.~\ref{app:gqw_calculations} for further information). In the following sections, we investigate this ability of QWs to find ground states, starting with the search problem---a well-studied toy problem that can be analytically solved using the QW.

\subsection{Quantum walk search}
\label{subsec:qw:search_problem}
In the search problem, we aim to find a specific bit string $z_{opt} \in \left\{0, 1\right\}^N$ from a set of $2^N$ possible strings. The problem can be mapped to a quantum setting using an oracle Hamiltonian
    \begin{align}
        \hat{H}_{O} = -\ket{z_{opt}}\bra{z_{opt}}
        \label{eq:H_Oracle}
	\end{align}
that assigns one unit of energy less to the solution state $\ket{z_{opt}}$ compared to the rest of basis states. Solving the search problem is then equivalent to finding the ground state of $\hat{H}_{O}$. The QW provides a means of solving the search problem by combining $\hat{H}_{O}$ with the driving Hamiltonian $\hat{H}_D$ (\equref{eq:H_D_Hypercube}), and adjusting their relative strength via the dimensionless hopping rate $\Gamma$. The computation is performed by evolving the quantum system, initialized in the equal superposition state
    \begin{align}
        \ket{\Psi_0} = \ket{+}^{\otimes N} = \frac{1}{\sqrt{2^N}} \sum_{j = 0}^{2^N} \,\ket{j},
        \label{eq:state_equal_superpos}
	\end{align}
under the QW search Hamiltonian 
    \begin{align}
        \label{eq:HQWS}
        \hat{H}_{QWS} = \Gamma \cdot \hat{H}_D + \hat{H}_O
    \end{align}
for a time $T$ and measuring the qubit register in the computational basis afterward.

Childs and Goldstone have solved the QW search problem analytically for various graph layouts~\cite{Childs2004CTQWSearchProblem}, including the complete and hypercube graphs. For each layout, they have calculated optimal values $\Gamma=\Gamma_{opt}$ (see also \cite{Childs2002QuantumSearchByMeasurement}) for which the performance of the QW search achieves the same optimal quadratic speed up as Grover's search algorithm~\cite{Grover1996Algorithm}.

\begin{figure}
  \centering
  \includegraphics[height=14cm]{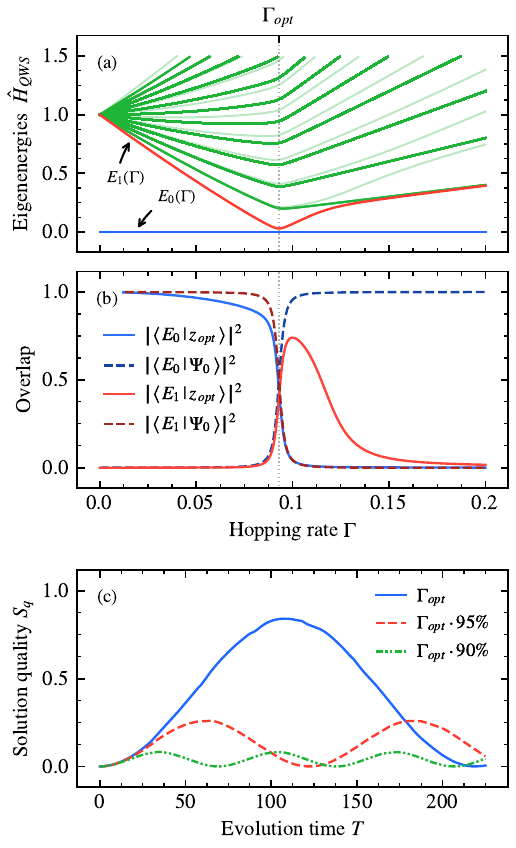}
  \caption{Application of the QW to an $N=12$ qubit search problem using the Hamiltonian $\hat{H}_{QWS}$ in Eq.~(\ref{eq:HQWS}). (a) Energy spectrum of $\hat{H}_{QWS}$ as a function of the hopping rate $\Gamma$. The blue and red curve denote the ground and first exited state, respectively. The data is shifted by the ground state energy. (b) Overlaps of the ground state (blue) and the first exited state (red) with the solution state (solid) and the initial state (dashed) as a function of $\Gamma$. (c) Solution quality $S_q$ as a function of the evolution time $T$ for $\Gamma = \Gamma_{opt}=1/2^N\sum_{r=1}^{N}\binom{N}{r}/2r$~\cite{Childs2004CTQWSearchProblem} (solid blue), $\Gamma = 95\% \cdot \Gamma_{opt}$ (dashed red) and $\Gamma = 90\% \cdot \Gamma_{opt}$ (dash-dotted green).}
  \label{fig:search_problem}
\end{figure}

Figure~\ref{fig:search_problem} presents the application of the QW to an $N=12$ qubit search problem, giving insights into the characteristics of the QW's dynamics. Specifically, the entire system evolves periodically, with the individual measurement probabilities of the vertices oscillating as a function of the evolution time $T$ (see~\figref{fig:search_problem}c). This behavior occurs because the QW performs Rabi oscillations between the initial state $\ket{\Psi_0}$ and the solution state $\ket{z_{opt}}$. Figure~\ref{fig:search_problem}a shows the two lowest energy levels corresponding to the states $\ket{E_0(\Gamma)}$ and $\ket{E_1(\Gamma)}$ as a function of the hopping rate $\Gamma$. When $\Gamma = \Gamma_{opt}$, the relative strengths of the two contributing Hamiltonians in $\hat{H}_{QWS}$ are balanced equally, and the two energy levels undergo an avoided level crossing. If $N$ is large enough, $\ket{E_{0, 1}(\Gamma_{opt})}$ is approximately equal to the uniform superposition of the initial and the solution state, i.e.~$\ket{E_{0, 1}(\Gamma_{opt})} \approx \left(\ket{\Psi_0} \pm \ket{z_{opt}}\right) / \sqrt{2}$ (see~\figref{fig:search_problem}b). Hence, $\hat{H}_{QWS}$ drives transitions between $\ket{\Psi_0}$ and $\ket{z_{opt}}$ with a frequency $\propto \left[E_1(\Gamma_{opt}) - E_0(\Gamma_{opt})\right]$. Consequently, the overlap with the solution state $|\braket{z_{opt} | \Psi\left(T\right)}|$ depends on the hopping rate $\Gamma$ and the evolution time $T$, which both require high precision in order to obtain accurate results (see.~\figref{fig:search_problem}c). 

\subsection{Guiding a quantum walk}
\label{subsec:qw:optimization_problem}
Combinatorial optimization problems can also benefit from the application of QWs. We can address optimization problems within the QW by augmenting the driver Hamiltonian $\hat{H}_D$ with the cost Hamiltonian $\hat{H}_C$ defined in \equref{eq:H_C}. The latter induces complex phase gradients between connected vertices based on their assigned cost value, thereby defining a direction of propagation for the walker in the graph. The QW optimization Hamiltonian on the hypercube mapping is given by
    \begin{align}
        \hat{H}_{QWO} = \Gamma \cdot \hat{H}_D + \hat{H}_C,
        \label{eq:H_QWO}
	\end{align}
with $\Gamma$ balancing the relative strength of the two contributing parts. The QW is performed analogously to \secref{subsec:qw:search_problem}, by initializing the qubits in the equal superposition state $\ket{\Psi_0}$ (see \equref{eq:state_equal_superpos}), evolving the system under $\hat{H}_{QWO}$ for a time $T$, and then measuring it in the computational basis.
\begin{figure}
  \centering
  \includegraphics[height=14cm]{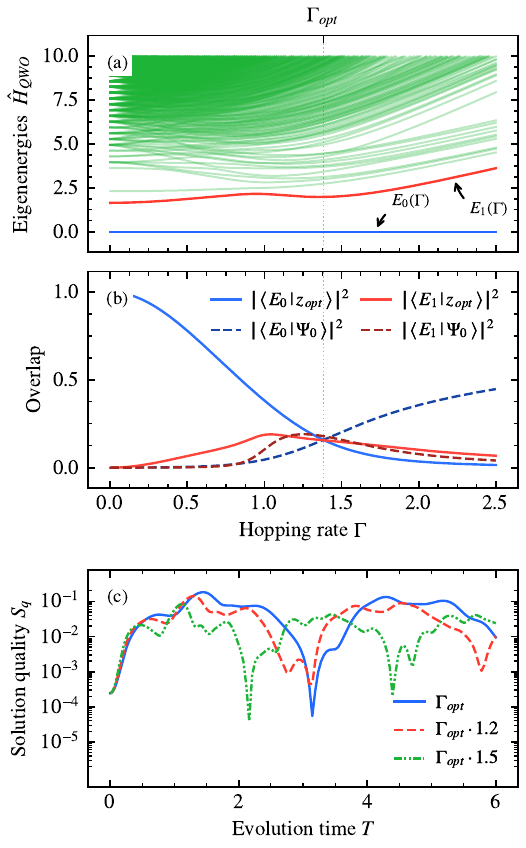}
  \caption{Application of the QW to solve an EC problem for $N=12$ qubits using the Hamiltonian $\hat{H}_{QWO}$ in Eq.~(\ref{eq:H_QWO}). (a) Energy spectrum of $\hat{H}_{QWO}$ as a function of the hopping rate $\Gamma$. The blue and red curve denote the ground and first exited state, respectively. The data is shifted by the ground state energy. (b) Overlaps of the ground state (blue) and the first exited state (red) with the solution state (solid) and the initial state (dashed) as a function of $\Gamma$. (c) Solution quality $S_q$ as a function of the evolution time $T$ for $\Gamma = \Gamma_{opt}$ (solid blue), $\Gamma = 1.2 \cdot \Gamma_{opt}$ (dashed red) and $\Gamma = 1.5 \cdot \Gamma_{opt}$ (dash-dotted green). The value of $\Gamma_{opt}$ has been computed numerically.}
  \label{fig:qw_exact_cover_problem}
\end{figure}
However, unlike for the search problem, the evolution under $\hat{H}_{QWO}$ cannot be efficiently calculated analytically. This makes it impractical to predict optimal parameter sets $\Gamma_{opt}$ and $T_{opt}$ which maximize the final overlap with the solution state for an arbitrary optimization problem. This is because the energy spectrum of $\hat{H}_C$ typically features numerous distinct levels with unknown energy gaps, in contrast to the almost completely degenerate spectrum of $\hat{H}_O$. Consequently, the energy levels of the combined Hamiltonian split as a function of $\Gamma$ and thereby undergo numerous avoided level crossings (see~\figref{fig:qw_exact_cover_problem}a). Since the system is initialized in a superposition across multiple energy levels, $\hat{H}_{QWO}$ drives transitions between various eigenstates, and the simple two-level description of the walker's dynamics used previously is no longer applicable (see~\figref{fig:qw_exact_cover_problem}b). As a result, the oscillation of the solution quality $S_q$ becomes highly complex, as multiple streams of amplitude transfers at different energy levels interfere with each other (see~\figref{fig:qw_exact_cover_problem}c).

Previous studies have explored heuristic approaches to obtain near-optimal hopping rates $\Gamma$ for the QW within polynomial time. For instance, Callison et al.~proposed estimating $\Gamma_{opt}$ from the overall energy scale of $\hat{H}_C$ by matching the total energy spreads of the two Hamiltonians in $\hat{H}_{QWO}$ \cite{FindingSpinglas_Callison}. Later, the authors extended this strategy by sampling $\Gamma_{opt}$ based on a maximization of the average dynamics on Sherrington–Kirkpatrick spin glass problems \cite{Callison2021EnergeticPerspectiveQAMultiStageQW}. Recently, Banks et al. investigated the link between time-independent Hamiltonians and thermalization, leading to an estimate of $\Gamma_{opt}$ through the eigenstate thermalization hypothesis on Max-Cut problem instances \cite{banks2023CTQWforMAXCUTarehot}.

While these strategies demonstrate the general ability of QWs to solve combinatorial optimization problems, they necessitate additional adjustments for each problem type (e.g., estimating energy gaps) and focus solely on the average dynamics in the hypercube graph. However, local variations in these dynamics across different regions of the graph are the primary reason for the observed distortions in the oscillation of the solution quality in \figref{fig:qw_exact_cover_problem}c. These distortions not only make it challenging to estimate $T_{\text{opt}}$ from a few samples, but also limit the maximally achievable solution quality for any $T$ as the system approaches a stationary state (cf.~Fig.~\ref{fig:IT_Sq_T_GQW-QW_N21} below). The optimal solution quality typically scales exponentially with the problem size $N$ (cf.~Fig.~\ref{fig:ec_Sq_T_GQW-QW-QA}b below) because the QW can only drive amplitude transfers within a fixed energy range, neglecting the amplitude originating from exponentially many states outside this range. Consequently, the QW as defined in Eq.~(\ref{eq:H_QWO}) is not well-suited for large combinatorial optimization problems.

\begin{figure}[H]
      \centering
      \includegraphics[]{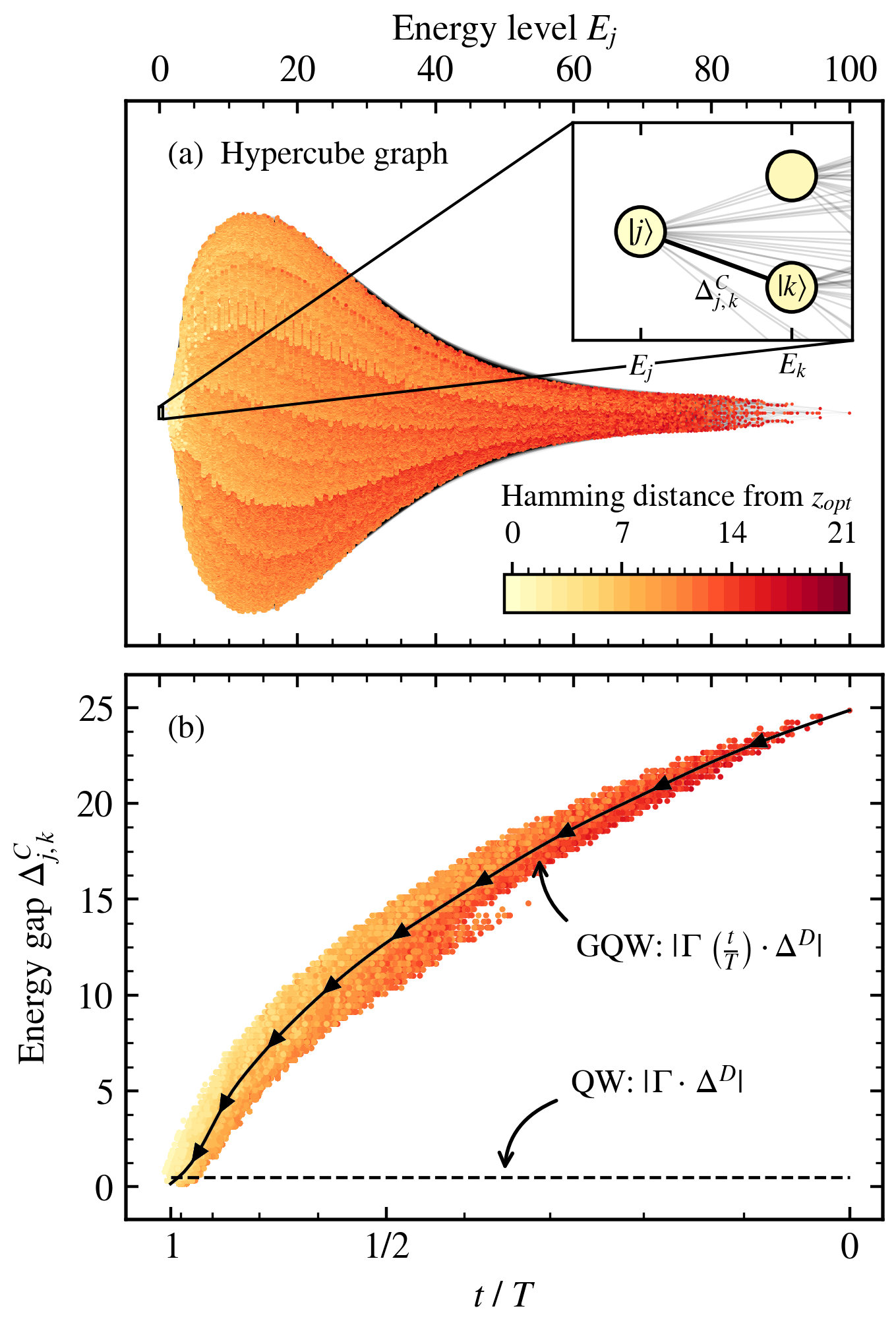}
      \caption{Energy spectrum analysis of an $N=21$ qubit EC problem. (a) Hypercube graph representation of $\hat{H}_C$, where each point corresponds to a computational basis state $\ket{j}$, ordered by increasing energy $E_j=\bra j \hat{H}_C\ket j$ from left to right. The color of each point represents the total Hamming distance to the ground state $z_{opt}$. Two states $\ket j$ and $\ket k$ are connected by an edge if the bitstrings $j$ and $k$ have a Hamming distance of $1$, meaning that the driver Hamiltonian $\hat{H}_D$ drives transitions between them. The inset in panel (a) provides a zoomed-in view near $z_{opt}$, highlighting the energy gaps $\Delta^C_{j,k}$ between connected states $\ket{j}$ and $\ket{k}$. (b) Distribution of the largest energy gaps $\Delta^C_{j,k}$ from a vertex $\ket{j}$ to a lower energy vertex $\ket{k}$ as a function of $E_j$. The data reveals an increasing trend of $\Delta^C_{j,k}$ with respect to $E_j$. The solid black curve represents a fit to the average energy gap $\braket{\Delta^C} (E)$ at energy level $E$ (using the scale of the top axis), used by the GQW to locally adjust the relative strength of the driving and problem Hamiltonian in Eq.~(\ref{eq:H_QWO}). The dashed black curve illustrates the optimal relative strength for a QW. The bottom axis denotes the relative time in both algorithms, with time progressing from right to left, as the GQW progresses from high-energy to low-energy amplitude transfers.}
      \label{fig:ec_gqw_graph}
    \end{figure}

Inspired by these strategies, but being interested in achieving practical quantum computation (as measured by a large solution quality) for any given evolution time $T$, problem size $N$ and problem type, we investigate the dynamics of QWs by applying the LAT theory to the hypercube graph $G$ shown in \figref{fig:ec_gqw_graph}a. By analyzing the transfer of probability amplitude in local subspaces spanned by basis-state pairs with Hamming distance 1 (i.e., states connected by an edge in $G$), we aim to derive a mechanism to control the movement of the walker locally in the graph, such that backpropagation of amplitude into (undesired) high-energy states can be suppressed. Thus, instead of maximizing transitions in the entire graph collectively, as proposed in prior study, our approach is to maximize them only locally at a time in order to guide the walker towards the solution state more effectively.

The LAT theory focuses on two-dimensional subspaces spanned by pairs of basis states, $\ket{j}$ and $\ket{k}$, that are connected in the hypercube graph $G$, i.e.~$\braket{j | \hat{H}_D | k} \neq 0$ (see inset of~\figref{fig:ec_gqw_graph}a). The effective two-level subspace Hamiltonian is given by
    \begin{align}
        \hat{H}_{QWO}^{(j,k)} = \Gamma \hat{H}_D^{(j, k)} + \hat{H}_C^{(j, k)},
	\end{align}
assuming the rest of the system remains in its initial state. Here, $\hat{H}_C^{(j, k)}$ denotes the local cost Hamiltonian and $\hat{H}_D^{(j, k)}$ is the local driving Hamiltonian, which are defined as
    \begin{align}
        \hat{H}_C^{(j, k)} = \begin{pmatrix}
                                \braket{j | \hat{H}_C | j} & 0\\
                                0 & \braket{k | \hat{H}_C | k}
                             \end{pmatrix} \qquad
        \hat{H}_D^{(j, k)} = -\begin{pmatrix}
                                0 & 1\\
                                1 & 0
                             \end{pmatrix}.
	\end{align}
If the system starts in the local equal superposition state, $\ket{+}^{(j, k)} = \left(\ket{j} + \ket{k}\right) / \sqrt{2}$, with the local energy gaps $\Gamma \Delta^D = -2 \Gamma$ of $\Gamma \hat{H}_D^{(j, k)}$ and $E_j - E_k = \Delta^C_{j, k} = 2 \delta$ of $\hat{H}_C^{(j, k)}$, the measurement probability of the desired lower-energy state $\ket{k}$, is given by
    \begin{align}
        P_{k}(t, \Gamma) = \frac{1}{2} + \frac{\Gamma \cdot \delta}{\Gamma^2 + \delta^2} \sin \left(\sqrt{\Gamma^2 + \delta^2} \cdot t\right)^2.
        \label{eq:P_Rabi}
	\end{align}
Equation~\eqref{eq:P_Rabi} represents a sinusoidal oscillation with a Rabi frequency of $\omega = \sqrt{\Gamma^2 + \delta^2}$ similar to~\secref{subsec:qw:search_problem} (note that this property holds for any initial state). Specifically, if $\Gamma \gg 1$, the driving Hamiltonian dominates $\hat{H}_{QWO}^{(j,k)}$, and the cost Hamiltonian has almost no influence on the system's evolution. Since $\ket{+}^{(j,k)}$ corresponds to the ground state of $\hat{H}_D^{(j,k)}$, the system remains primarily in its initial state. Conversely, if $\Gamma \ll 1$, $\hat{H}_C^{(j,k)}$ governs the evolution. Since $\hat{H}_C^{(j,k)}$ is diagonal and only induces phase rotations in the computational basis, no amplitude transfer occurs. Only for $\vert\Gamma \Delta^D \vert \approx \Delta^C_{j, k}$, the two Hamiltonians' relative strengths are balanced, and transitions in the local subspace are maximized.

\bigbreak
   
As a result, amplitude transfers can only occur efficiently among specific subsets of vertex pairs with an energy gap $\Delta^C_{j, k} \approx \vert \Gamma \Delta^D \vert$ in $\hat{H}_C$. Transitions between states with significantly larger or smaller energy gaps are suppressed for fixed $\Gamma$. Since the energy gaps of vertex pairs typically vary throughout the graph, we can steer the walker's movement effectively by selecting $\Gamma$ to activate only the desired transitions in the graph. 

For all EC, TSP, and GO problems under investigation, we noticed empirically that the distribution of the energy levels acquires a characteristic 'onion shape' (see~\figref{fig:ec_gqw_graph}a) as soon as the problem instance is sufficiently complex (i.e., the number of qubits is sufficiently large). Additionally, the largest energy gap $\Delta^C_{j, k}$ from vertex $\ket{j}$ to a lower energy vertex $\ket{k}$ increases approximately monotonically as a function of its energy level $E_j$ (see~\figref{fig:ec_gqw_graph}b).
Consequently, large values of $\Gamma$ are usually optimal in the regime of high-energy states, while small values are preferable near the solution state. Choosing a fixed value for $\Gamma$ has the disadvantage of limiting the maximally achievable success probability because not all edges can sufficiently contribute to the amplitude transport. In particular, only amplitude transfers within a fixed energy range can be addressed, generally prohibiting states located at high energies from transporting amplitude to the solution state.

The strategy we propose in this paper is based on an energy-dependent hopping rate, where $\vert \tilde{\Gamma}(E) \cdot \Delta^D \vert$ corresponds to the average $\braket{\Delta^C}\left(E\right)$ of the largest energy gaps of $\hat{H}_C$ at energy level $E$ in the graph. The approach is to confine the walker to a gradually shrinking energy region around the solution state by progressing monotonically from high-energy to low-energy optimal hopping rates $\tilde{\Gamma}$. Therefore, we set $\Gamma(t) = \tilde{\Gamma}\left(E\left(t\right)\right)$, where $E\left(t\right)$ is a monotonic sweep, and define
    \begin{align}
        \hat{H}_{GQW} = \Gamma(t) \cdot \hat{H}_D + \hat{H}_C.
        \label{eq:H_GQW}
	\end{align}
Equation~\ref{eq:P_Rabi} shows that the frequency of the local Rabi oscillations depends on the size of the energy gaps, resulting in faster amplitude transfers at larger gaps (see \figref{fig:ec_gqw_graph}b). Thus, the GQW needs to spend less time driving transitions at high energies than at small energies to avoid amplitude going back to high energy state. Consequently, the energy sweep must be rescaled according to $\braket{\Delta^C}(E)$, yielding the rescaled time
    \begin{align}
        \label{eq:dEdt}
        s(t) = \frac{\int_{E_{max} + (E_{min}-E_{max}) \cdot 
        t/T}^{E_{max}} \braket{\Delta^C}(E) \; dE}{ \int_{E_{min}}^{E_{max}} \braket{\Delta^C}(E) \; dE}
    \end{align}
with $E(t) = E_{max} + (E_{min}-E_{max}) \cdot s(t)$, where $E_{min}$ and $E_{max}$ denote the lowest and highest energy levels of $\hat{H}_C$, respectively, and $T$ is the total evolution time. At $t=0$, the algorithm starts with a relatively high hopping rate ($\Gamma = \tilde{\Gamma}(E_{max})$) that maximizes amplitude transfers only at high energy levels and suppresses dynamics at low and intermediate energies. As the hopping rate is decreased over time, the system starts to perform amplitude transfers at lower energy levels. Simultaneously, subspaces at higher energies become gradually detuned again, suppressing the action of the driving Hamiltonian. This is essential as it prevents the backpropagation of probability into high-energy states and enables us to actively guide the walker towards low-energy states.

The primary advantage of this approach lies in the establishment of a continuous amplitude flow from all computational basis states towards the solution state, consequently overcoming the previous limitation on the maximally achievable solution quality. Furthermore, the latter is no longer subject to the complex oscillations, which required a precisely chosen evolution time $T$ (cf.~\figref{fig:qw_exact_cover_problem}c). Instead, $S_q(T)$ follows a collective sinusoidal oscillation of \equref{eq:P_Rabi}, where the selection of $\Gamma(t)$ ensures that active subspaces oscillate with similar Rabi frequencies while high energy subspaces become gradually detuned, hence suppressing the backpropagation of probability amplitude. By varying the evolution time $T$, the speed at which $\tilde{\Gamma}(E)$ is swept can be adjusted, thereby altering the duration that groups of subspaces remain active. As a result, $S_q(T)$ depends solely on the order of magnitude of $T$, rather than its precise value.

To evaluate the efficiency of a GQW, we compare its performance to a conventional QW on an EC problem instance comprising $N=21$ qubits. Figure~\ref{fig:ec_gqw_graph}b shows the average $\braket{\Delta^C} (E)$ of the largest energy gaps between connected vertices at each energy level $E$. We obtain $\vert \tilde{\Gamma}(E) \cdot \Delta^D \vert$ by fitting a polynomial of degree $6$ to the data points (see the black arrows in~\figref{fig:ec_gqw_graph}b). The bottom axis indicates the non-linear sweep of $E\left(t\right)$. In comparison, for the QW, we determined $\Gamma_{opt}$ by sampling a classical optimizer (see dashed black curve). $\Gamma_{opt}$ corresponds to the optimal hopping rate that yields the highest success probability over $T \in \left[0.1, 10.0\right]$.

\begin{figure}
  \centering
  \includegraphics[]{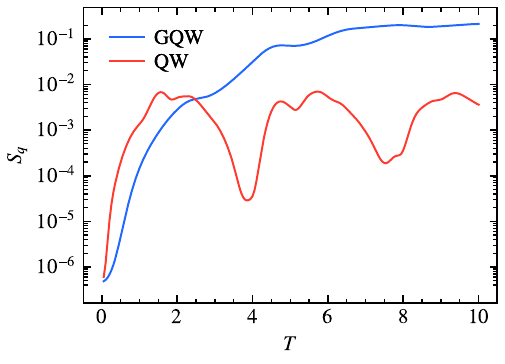}
  \caption{Performance comparison between the GQW (blue) and the QW (red) on the $N=21$ qubit EC problem depicted in \figref{fig:ec_gqw_graph}. The GQW dynamically adjusts the relative strength $\Gamma$ of the driving and problem Hamiltonian in \equref{eq:H_GQW} based on the average $\braket{\Delta^C} (E)$ of the largest energy gaps in $\hat{H}_C$. In contrast, the QW employs a fixed hopping rate $\Gamma$ that has been pre-optimized (see~\figref{fig:ec_gqw_graph}b). The performance is evaluated using the solution quality $S_q$, as defined in \equref{eq:S}. Note that the results presented here are indicative and require knowledge of the complete energy spectrum.}
  \label{fig:IT_Sq_T_GQW-QW_N21}
\end{figure}

Figure~\ref{fig:IT_Sq_T_GQW-QW_N21} compares the performance of the GQW and the QW as a function of the total evolution time $T$. For $T \leq 2$, both strategies exhibit similar behavior, showing a rapid increase in solution quality, with the QW achieving its peak at $T=1.5$. Notably, for $T\leq2.5$, the GQW yields lower solution qualities compared to the QW. This discrepancy arises from the relatively small optimal hopping rate of the QW, enabling it to focus its evolution on amplitude transitions near and into the solution state. In contrast, the GQW considers the entire graph and therefore spends the initial part of its evolution at high and intermediate energy levels (the right part of \figref{fig:ec_gqw_graph}b). This approach leads to very short evolutions at each energy level for small $T$, causing only fractions of the probability amplitudes to be transported towards $z_{opt}$. However, as $T$ increases, the situation changes, and the GQW obtains superior solution qualities to the QW for $T\geq2.5$. Here, the QW exhibits oscillatory behavior between $10^{-5}$ to $10^{-2}$. In contrast, the GQW follows a monotonic increase, where $S_{q}(T)$ saturates at approximately $21\%$, outperforming the QW by approximately one order of magnitude.

The observed saturation results from imbalances within the local subspaces at small $E$. As a large amount of amplitude accumulates in the ground state, the system progressively deviates from the state of equal local superposition. Consequently, driving these subspaces with their respective optimal hopping rate $\Tilde{\Gamma}(E)$ eventually redirects amplitude back into the higher energy state, thus limiting $S_{q}(T)$ (see also App.~\ref{app:gqw_calculations}).

\subsection{Practical guided quantum walk}
\label{subsec:qw:guided_quantum_walk}
The previous section has shown the potential of GQWs for solving combinatorial optimization problems, by adjusting the relative strength of the two Hamiltonians in $\hat{H}_{GQW}$ based on the walker's position in the graph. The main benefits are an increased maximum solution quality and a suppression of complex oscillations in $S_{q}(t)$, providing good results for arbitrary $T$ and $N$.

Of course, obtaining the optimal function $\tilde{\Gamma}\left(E(t)\right)$ is generally not efficiently possible, because it requires knowledge of the entire energy spectrum of $\hat{H}_C$. In order to still make use of the promising methodology described above, we propose a variational ansatz, in analogy to other quantum optimization algorithms~\cite{Farhi2014QAOA,McClean2016TheoryVariationalQuantumEigensolver,Tilly20221VQE}. The idea is to imitate
the optimal distribution $\tilde{\Gamma}\left(E\right)$ by a function $\tilde{\Gamma}\left(E, \boldsymbol{\lambda}\right)$, which is tuned using a set of $M$ hyperparameters $\boldsymbol{\lambda} = \left(\lambda_1, \dots, \lambda_M \right)$. Note that henceforth we consider only linear sweeps of the energy spectrum, i.e.~$E\left(t\right) = \left(E_{min} - E_{max}\right) \cdot t/T + E_{max}$, thus encoding the sampling speed in the shape of $\tilde{\Gamma}$. The hope is that as long as $\tilde{\Gamma}\left(E, \boldsymbol{\lambda}\right)$ describes $\tilde{\Gamma}\left(E\right)$ closely enough, similar dynamics to~\secref{subsec:qw:optimization_problem} can be obtained. In fact, introducing hyperparameters into the algorithm even enables the guided quantum walk to overcome its limitations at small and large $T$. For instance, at small $T$, the GQW could selectively model $\tilde{\Gamma}\left(E\right)$ only up to $E < E_{max}$, thereby operating solely on a subspace of the entire graph near $\ket{z_{opt}}$. Conversely, at large $T$, the GQW can prevent the transfer of amplitude into higher energy states when operating at small $E$ by compensating the imbalances within the local subspaces through a reduction of the final hopping rate, $\tilde{\Gamma}\left(E = E_{min}\right)$.

The proposed algorithm employs a hybrid quantum-classical ansatz, in which a classical optimizer adjusts the set of hyperparameters $\boldsymbol{\lambda}$ based on the minimization of the energy expectation value $E_{\Psi}$ (see \equref{eq:<H_C>}) of the final quantum state $\ket{\Psi}$ obtained by a quantum device performing the GQW. Note that the number $M$ of hyperparameters is fixed, and we investigate the impact of this optimization phase on the total run time in~\secref{subsec:results:optimization_complexity}. 

We propose a function $\tilde{\Gamma}\left(E, \boldsymbol{\lambda}\right)$ based on cubic Bézier curves. We chose Bézier curves instead of simple polynomials because we expect the optimal hopping rate to be smooth and monotonically decreasing in $E$. Although polynomials can produce such functions for $E \in \left[E_{min}, E_{max}\right]$, their parameters are generally hard to tune, as small changes can lead to substantially different functions. In contrast, Bézier curves are much easier to optimize since their general shape can be predetermined. Moreover, the latter varies continuously and sufficiently slowly in its parameters, resulting in a sufficiently smooth search space for $\boldsymbol{\lambda}$. Due to these properties, we strongly encourage their use in other fields of quantum computing, such as optimizing annealing schedules \cite{QAOpt_PathOptZeng, QAOpt_PathOptHerr, QAOpt_MonteCarloTreeSearch} or deriving optimal parameter sets for the quantum approximate optimization algorithm (QAOA) \cite{Farhi2000AdiabaticQuantumComputation, Farhi2014QAOA, QAOA_Opt_Zhou, QAOA_Opt_Sack, SC_Willsch}.

\begin{figure}
      \centering
      \includegraphics[]{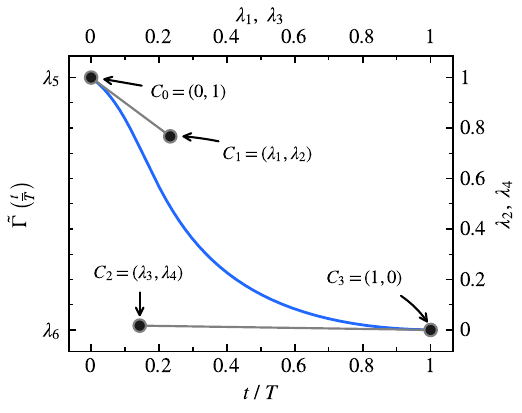}
      \caption{Optimal hopping rate $\tilde{\Gamma}\left(t, \boldsymbol{\lambda}\right)$ (blue) for the $N=21$ qubit EC problem shown in \figref{fig:ec_gqw_graph}, with a total evolution time of $T=10$. The hopping rate is represented by a cubic Bézier curve (see \equref{eq:Bezier_curves}) and is defined by four control points $C_{0,1,2,3}$ based on the hyperparameters $\lambda_{1-4}$, along with the boundary conditions $\tilde{\Gamma}\left(t=0, \boldsymbol{\lambda}\right) = 10^{2 \cdot \lambda_5}$ and $\tilde{\Gamma}\left(t=T, \boldsymbol{\lambda}\right) = 10^{-3 \cdot \lambda_6}$. The curve is obtained by optimizing a set of six hyperparameters $\boldsymbol{\lambda}$ using the classical Nelder-Mead optimizer~\cite{NelderMead1965,numericalrecipes}, employing $N_{opt} = 100$ optimization steps starting from a random initial configuration.}
      \label{fig:EC_Hyperparam_Gamma}
    \end{figure}

Cubic Bézier curves are based on Bernstein polynomials and are defined through four control points $C_i = \left(x_i, y_i\right)^T$ in a two-dimensional plane as
    \begin{align}
        x(\tau) = x_0\left(1 - \tau\right)^3 &+ 3x_1\left(1 - \tau\right)^2 \tau \nonumber\\&+ 3x_2\left(1 - \tau\right)\tau^2 + x_3\tau^3 \\
        y(\tau) = y_0\left(1 - \tau\right)^3 &+ 3y_1\left(1 - \tau\right)^2 \tau \nonumber\\&+ 3y_2\left(1 - \tau\right)\tau^2 + y_3\tau^3.
        \label{eq:Bezier_curves}
	\end{align}
Choosing $x_0 < x_1,x_2 < x_3$, the curve can be mapped into one dimension, by solving $t = x(\tau)$ for $\tau(t)$ and inserting it into $y$. Based on the observations made in~\secref{subsec:qw:optimization_problem}, we set $C_0 = (0, 1)$ and $C_3 = (1, 0)$ to ensure a monotonically decreasing function, with $4$ hyperparameters $\lambda_{1, 2, 3, 4} \in \left(0, 1\right)$ controlling its shape, i.e.~$C_1 = \left(\lambda_1, \lambda_2\right)^T$ and $C_2 = \left(\lambda_3, \lambda_4\right)^T$. The hopping rate is then given by $\tilde{\Gamma}\left(t, \boldsymbol{\lambda}\right) = y(t) \, \cdot \, 10^{2 \cdot \lambda_5} \, + \, \left(1 - y(t)\right) \, \cdot \, 10^{-3 \cdot \lambda_6}$, where we introduced two additional hyperparameters, $\lambda_5, \lambda_6 \in \left[0, 1\right]$, that describe the boundary conditions $\tilde{\Gamma}\left(t=0, \boldsymbol{\lambda}\right) = 10^{2 \cdot \lambda_5}$ and $\tilde{\Gamma}\left(t=T, \boldsymbol{\lambda}\right) = 10^{-3 \cdot \lambda_6}$. $\lambda_5$ determines the energy range in which the GQW operates in the graph (see discussion for small $T$) and $\lambda_6$ controls the detuning of the local subspaces near the solution state (see discussion for large $T$). Since the optimal scale of $\tilde{\Gamma}\left(t, \boldsymbol{\lambda}\right)$ can vary significantly between different problem instances, we chose an exponential rescaling to simplify the optimization landscape. Note that the parameters $-2$ and $3$ were found suitable for the problems under investigation, but generally depend on the energy scale of the cost Hamiltonian $\hat{H}_C$. In \figref{fig:EC_Hyperparam_Gamma} we present $\tilde{\Gamma}\left(t, \boldsymbol{\lambda}\right)$ for the aforementioned EC problem and $T=10$.
    
\subsection{From quantum walk to quantum annealing}
\label{subsec:qw:relation_qa}
The LAT theory, used in the derivation of the GQW (see \secref{subsec:qw:optimization_problem}), offers a new perspective on the working principle of quantum annealing (QA) and its relationship to QWs. QA belongs to the class of continuous-time quantum optimization algorithms that rely on an adiabatic transition from the driver Hamiltonian $\hat{H}_D$ (i.e., $\Gamma\left(t=0\right) \rightarrow \infty$) to the problem Hamiltonian $\hat{H}_C$ (i.e., $\Gamma\left(t=T\right) = 0$) throughout the time evolution. According to the adiabatic theorem of quantum mechanics, if this transition occurs sufficiently slowly, and the system is initially prepared in the ground state of $\hat{H}_D$, it will remain in the instantaneous ground state of the combined Hamiltonian $\hat{H}_{GQW}$ in Eq.~\eqref{eq:H_GQW}, ultimately reaching the solution state $\ket{z_{opt}}$.

    \begin{figure}
      \centering
      \includegraphics[width=\columnwidth]{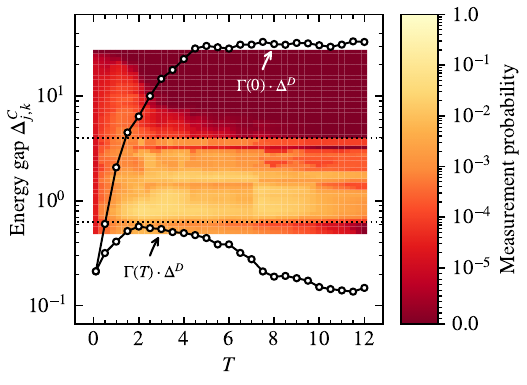}
      \caption{Ranges of the optimal hopping rate $\tilde{\Gamma}\left(t, \boldsymbol{\lambda}\right)$ (solid lines) for an $N=15$ qubit EC problem as a function of the total evolution time $T$. $\tilde{\Gamma}\left(t=0, \boldsymbol{\lambda}\right)$ and $\tilde{\Gamma}\left(t=T, \boldsymbol{\lambda}\right)$ denote the initial and final hopping rate, respectively, as determined by the hyperparameters $\lambda_5$ and $\lambda_6$ (see \secref{subsec:qw:guided_quantum_walk}). Note that $\tilde{\Gamma}\left(t, \boldsymbol{\lambda}\right)$ is scaled by $\Delta^D$ to map to the energy gaps of the local subspaces in the hypercube graph. The heat map presents the evolution of the measurement probabilities averaged over states with equal maximum energy gaps $\Delta^C_{j,k}$ (see \figref{fig:ec_gqw_graph}b)). Dotted lines mark the range of energy gaps of local subspaces involving the solution state.}
      \label{fig:ec_Params_Evo}
    \end{figure}

\begin{figure*}
      \centering
      \includegraphics[]{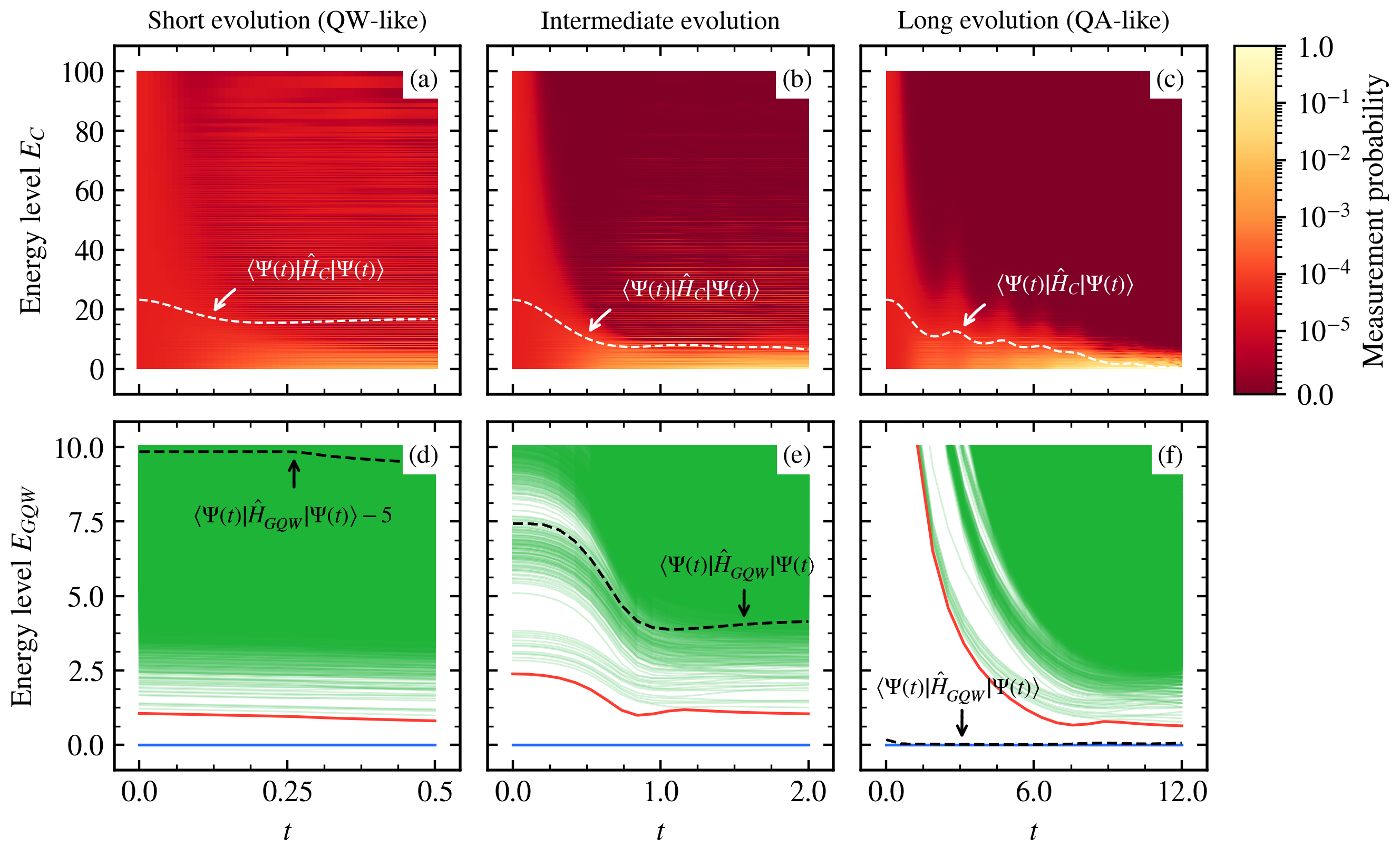}
      \caption{Optimized quantum evolutions performed by the GQW on the $N=15$ qubit EC problem shown in \figref{fig:ec_Params_Evo} for short ($T=0.5$, (a) and (d)), intermediate ($T=2.0$, (b) and (e)), and long ($T=12$, (c) and (f)) evolution times. Panels (a)--(c) illustrate the average measurement probabilities of the energy levels $E_C$ of the problem Hamiltonian $\hat{H}_C$. Panels (d)--(f) display the evolution of the instantaneous energy levels $E_{GQW}$ of the combined Hamiltonian $\hat{H}_{GQW}$ (\equref{eq:H_GQW}). The blue and red curves represent the instantaneous ground state and the first excited state, respectively. Dashed lines are included in all panels to indicate the instantaneous energy expectation values of the system with respect to $\hat{H}_C$ (panels (a)--(c)) and $\hat{H}_{GQW}$ (panels (d)--(f)). The final solution qualities are $S_q = 0.7\%$ ($T=0.5$), $S_q = 13.6\%$ ($T=2.0$), and $S_q = 91.3\%$ ($T=12.0$).}
      \label{fig:ec_evo_T}
    \end{figure*}

To explore the relationship between QWs and QA, we use the GQW to examine the strategies governing optimal quantum evolutions across different time intervals $T$. Figure~\ref{fig:ec_evo_T} presents simulation results of the GQW applied to an $N=15$ qubit EC problem, covering short ($T=0.5$), intermediate ($T=2.0$), and long ($T=12$) evolutions. Panels (a)--(c) depict the average measurement probabilities of the energy levels $E_C$ within the problem Hamiltonian $\hat{H}_C$. Panels (d)--(f) showcase the evolution of the instantaneous energy levels $E_{GQW}$ of the combined Hamiltonian $\hat{H}_{GQW}$. Furthermore, \figref{fig:ec_Params_Evo} presents the optimal range of $\tilde{\Gamma}\left(t, \boldsymbol{\lambda}\right)$ as a function of the total evolution time $T$ for the same EC problem.

\bigbreak

\paragraph*{Short evolutions:} In the case of short evolutions, the optimal hopping rate schedule $\tilde{\Gamma}\left(t, \boldsymbol{\lambda}\right)$ determined by the GQW maintains a near-constant relationship between $\hat{H}_D$ and $\hat{H}_C$ in \equref{eq:H_GQW}. Consequently, the instantaneous energy levels mostly remain unchanged during the evolution, and the system starts and ends in a superposition of the instantaneous basis states (see the dashed line in \figref{fig:ec_evo_T}d). For $T = 0.5$, $\tilde{\Gamma}\left(t, \boldsymbol{\lambda}\right)$ decreases from $0.6$, approximately equivalent to $\Tilde{\Gamma}\left(E_{min}\right)$, to $0.3$. This transition initially drives the subspaces involving $\ket{z_{opt}}$, hence resembling a QW-like procedure, before progressively detuning them to prevent the back-propagation of amplitude into higher energy states (see \figref{fig:ec_Params_Evo}). The latter is necessary as these subspaces gradually move away from a close-to-equal superposition state (see App.~\ref{app:gqw_calculations}). Figure~\ref{fig:ec_evo_T}a demonstrates that this strategy results in a guided movement of amplitude towards $\ket{z_{opt}}$ for $E_C \leq 12$, evident from the emerging gradient in the measurement probabilities. Probability amplitude originating from intermediate and high energy levels, on the other hand, fails to reach the solution state and instead becomes trapped at these energies within the graph (see horizontal dark and bright stripes in \figref{fig:ec_evo_T}a). This highlights the inherent limitations of QWs.

\bigbreak

\paragraph*{Intermediate evolutions:} In the case of intermediate evolution times, the optimal hopping rate schedule operates over a wider range of values, with the initial hopping rate $\tilde{\Gamma}\left(t=0, \boldsymbol{\lambda}\right)$ increasing monotonically as a function of $T$ (see \figref{fig:ec_Params_Evo}). In doing so, the GQW extends the subset of the hypercube graph in which amplitude is actively guided through the local subspaces. This is evident from the increased final success probability and the gradual transport of amplitude from high to low energy states for $t \leq 0.5$ (see \figref{fig:ec_evo_T}b). The presence of bright stripes at $E_C \geq 10$ indicates that the GQW is still operating on a subset of the hypercube graph, causing amplitude from high-energy states to become trapped at intermediate energy levels. This demonstrates that even for intermediate evolution times, it is more advantageous to neglect amplitude at high-energy states and focus on amplitude transfers near the solution state. Interestingly, also the final hopping rate $\tilde{\Gamma}\left(t=T, \boldsymbol{\lambda}\right)$ increases for longer evolution times $T$ up to a maximum at $T \approx 2$. Subsequently, for $T > 2$, $\tilde{\Gamma}\left(t=T, \boldsymbol{\lambda}\right)$ declines, again detuning the local subspaces at small $E$. The shape of $\tilde{\Gamma}\left(t=T, \boldsymbol{\lambda}\right)$ is likely influenced by the density of states of $\hat{H}_C$. As \figref{fig:ec_gqw_graph}a) illustrates for an $N=21$ qubit EC problem, the state density of $\hat{H}_C$ for our EC instances rapidly increases for smaller $E$, reaching a peak at $E_p$, followed by an exponential decline. Consequently, as $\tilde{\Gamma}\left(t=0, \boldsymbol{\lambda}\right)$ initially increases, the accessible amplitude likely grows faster than the transport of amplitude into $\ket{z{opt}}$ within $T$. This keeps the local subspaces surrounding $\ket{z_{opt}}$ closer to an equal superposition state, requiring less final detuning. The peak of $\tilde{\Gamma}\left(t=T, \boldsymbol{\lambda}\right)$ aligns with the point where $\tilde{\Gamma}\left(t=0, \boldsymbol{\lambda}\right) \approx \Tilde{\Gamma}(\left(E_p\right)$. As $\tilde{\Gamma}\left(t=0, \boldsymbol{\lambda}\right)$ further increases, the growth of accessible amplitude slows down, causing the local subspaces at lower $E$ to move away from an equal superposition state. Consequently, a larger final detuning is required, resulting in a smaller $\tilde{\Gamma}\left(t=T, \boldsymbol{\lambda}\right)$.

\bigbreak

\paragraph*{Long evolutions:} For long evolutions, the optimal hopping rate schedule determined by the GQW resembles a QA-like schedule by transitioning nearly entirely from $\hat{H}_D$ to $\hat{H}_C$ in \equref{eq:H_GQW} (see \figref{fig:ec_Params_Evo}). Consequently, the system follows the instantaneous ground state of $\hat{H}_{GQW}$ throughout the evolution, indicated by the dashed line in \figref{fig:ec_evo_T}f. Throughout this process, the GQW effectively drives amplitude transfers across the entire graph, initiating from high energy levels and moving towards low energies, evident by the absence of horizontal stripes in \figref{fig:ec_gqw_graph}c. Notably, the confinement of amplitude occurs exponentially in $t$, indicating a non-linear sweep through the energy spectrum (cf. \equref{eq:dEdt}). Furthermore, the propagation of amplitude follows a wave-like pattern, as illustrated by the dashed line in \figref{fig:ec_evo_T}c. This pattern arises due to the confinement of amplitude in a decreasing number of vertices, leading to deviations from the equal superposition states in the local subspaces. This results in a temporary backpropagation of amplitude into higher energy states (see App.~\ref{app:gqw_calculations}). Nevertheless, the continuous decrease of the hopping rate $\tilde{\Gamma}\left(t, \boldsymbol{\lambda}\right)$ ensures, on average, the transportation of amplitude into the solution state. 

\bigbreak

The hopping rate schedules derived from the GQW not only emphasize the intrinsic connection between QWs and QA but also highlight the existence of optimal quantum evolutions that extend beyond the scopes of these two algorithms. While a QW-like strategy proves optimal for short evolutions and a QA-like procedure is favored for long evolutions, our investigations reveal that intermediate values of $T$ necessitate a combination of both strategies to maximize the solution quality.

This observation can be explained through LAT theory (see \secref{subsec:qw:optimization_problem} and App.~\ref{app:gqw_calculations}), as QWs and QA can be viewed as two distinct formulations of the same underlying concept. Both approaches aim for the optimal transfer of probability amplitude within local subspaces of the graph. QWs achieve this by employing a constant, small hopping rate $\Gamma$, focusing exclusively on direct transfers into the solution state during short evolutions. However, this strategy becomes less effective for long evolutions, where sufficient time is available to guide amplitude at higher energy levels as well. Consequently, QA guides amplitude through the entire graph by linking multiple local QWs together, employing a continuously decreasing hopping rate $\tilde{\Gamma}\left(t\right)$. Thus, QWs and QA represent the two extremes within the GQW framework, with one concentrating solely on subspaces around the solution state (i.e., $\tilde{\Gamma}\left(t=0\right) \approx \tilde{\Gamma}\left(t=T\right) \ll 1$) and the other considering the entire graph (i.e., $\tilde{\Gamma}\left(t=0\right) \gg 1$ and $\tilde{\Gamma}\left(t=T\right) \ll 1$, cf.~Fig.~\ref{fig:EC_Hyperparam_Gamma}).

The GQW operates in the transition region between these two extremes, striking a balance between the number of guided local subspaces (i.e., the amount of guided amplitude) and the time spent at each energy level (i.e., the amount of amplitude transferred within each local subspace) for a given $T$. As will be discussed in \secref{subsec:results:scaling_N}, these intermediate evolutions, which surpass the limitations of adiabatic time evolutions, might be capable of effectively solving large-scale optimization problems.


\begin{figure*}
    \centering
    \includegraphics[]{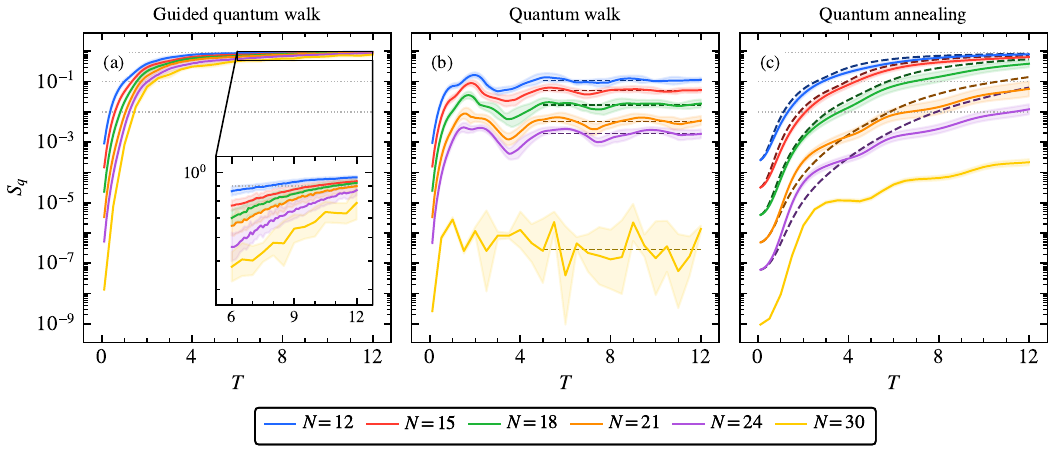}
    \caption{Performance comparison of (a) the GQW, (b) the QW, and (c) QA as a function of the total evolution time $T$ on EC problems with problem sizes $N \in \left\{12, 15, 18, 21, 24, 30\right\}$ (see legend). Panel (c) presents simulation results for both linear QA (solid lines) and optimized locally adiabatic QA~\cite{LQA_Roland} (dashed lines). The latter uses a rescaled time based on the size of the instantaneous energygap during the annealing process, assuming complete knowledge about the optimization problem.} The performance of all algorithms is assessed based on the solution quality $S_q$ defined in \eqref{eq:S}. Each problem size $N$ is evaluated using a set of $10$ randomly generated problem instances, and the presented data corresponds to the geometric mean solution qualities (solid curves) along with the geometric standard deviations (shaded areas). Note that the GQW and the QW undergo a prior classical optimization phase. The dotted gray lines indicate horizontal cuts ($S_q \in \left\{1\%, 10\%, 90\%\right\}$), which are detailed in \figref{fig:ec_T_N_GQW-QA}. The inset in panel (a) provides a zoomed-in view of the region corresponding to large evolution times $T$. Additionally, the dashed lines in panel (b) indicate the stationary solution quality that the damped oscillation of $S_q(T)$ approaches for the QW at large $T$.
    \label{fig:ec_Sq_T_GQW-QW-QA}
\end{figure*}

\section{Results}
\label{sec:results}
In this section, we present a comprehensive performance analysis of the GQW. We compare the solution quality $S_q$ both as a function of the evolution time $T$ and the system size $N$ to the conventional QW and QA (see \secref{subsec:results:comparison}). Section~\ref{subsec:results:scaling_N} provides scaling results of the three algorithms in the regime of large problem instances (i.e., $N \gg 1$). Our analysis indicates that the GQW achieves significantly higher solution qualities within a linearly growing timespan, rendering it a promising candidate for near-term quantum devices. Finally, in \secref{subsec:results:optimization_complexity} we address the impact of the classical optimization phase on the total run time, demonstrating that the average time-to-solution scales better by a factor of $\approx2$ ($\approx4$) compared to linear QA (QW). 

\subsection{Comparison of GQW, QW and QA}
\label{subsec:results:comparison}
We assess the effectiveness of the GQW through numerical simulations carried out on the JUWELS Booster supercomputer at the J\"ulich Supercomputing Centre of the Forschungszentrum J\"ulich~\cite{JuwelsClusterBooster}. We compare its performance against a conventional QW and QA. Our simulations consider EC and GO instances with problem sizes $N \in \left\{12, 15, 18, 21, 24, 30\right\}$, as well as TSP instances with $N \in \left\{9, 16, 25\right\}$ qubits. To demonstrate the generality of our findings, we examine $10$ randomly generated problem instances for each problem type and size. We remark that the energy spectrum of each problem has been obtained classically, providing the total energy range and the individual energies of the valid states for the calculation of $S_q$ (see \equref{eq:S}). Note, however, that this is done only for the purpose of benchmarking, and it is not required to apply the GQW in a practical scenario. 

To obtain the final quantum state of the system at the end of each algorithm, we use the second-order Suzuki-Trotter product formula algorithm~\cite{SUZUKI1993,Raedt2006} to solve the time-dependent Schrödinger equation, $i \, \partial t \ket{\Psi(t)} = H(t) \ket{\Psi(t)}$, with a time step of $10^{-5}$ and total evolution times $T$ ranging between $0.1$ and $12.0$. Given the consistency of our findings, we present the results for the EC instances in \figref{fig:ec_Sq_T_GQW-QW-QA}, showing the obtained solution qualities averaged within each system size, along with the standard deviations. The analogous results for the TSP and GO problems are available in App.~\ref{app:go_tsp_results} and support the same conclusions.

\bigbreak

\paragraph*{Guided quantum walk:}
The GQW is optimized for each problem instance and evolution time $T$ individually by tuning its $6$ hyperparameters $\boldsymbol{\lambda}$ to minimize the energy expectation value $E_{\Psi}$ (see \equref{eq:<H_C>}) of the final quantum state $\ket{\Psi}$. Each parameter set is thereby selected from a pool of $N_{rep} = 100$ repetitions of the Nelder-Mead classical optimizer, where in each sample the $6$ parameters are initialized randomly and adjusted a maximum of $N_{opt} = 100$ times. We choose this approach to ensure that the algorithm converges to a near-optimal minimum in the parameter search space. However, we note that a sufficient set of parameters is typically found within the first $N_{rep} = 20$ repetitions. In \secref{subsec:results:optimization_complexity}, we discuss the impact of this optimization phase on the total run time.

Figure~\ref{fig:ec_Sq_T_GQW-QW-QA}a presents the simulation results of the GQW, showing the scaling of the solution quality $S_q(T)$ as a function of the evolution time $T$. Across various problem types and sizes, we observe consistent patterns in the scaling of $S_q(T)$, corresponding to the three regimes of evolutions (see \secref{subsec:qw:relation_qa}).

Initially, the solution quality exhibits rapid growth, matching the results obtained by the QW algorithm for $T \leq 0.5$ (see \figref{fig:ec_Sq_T_GQW-QW-QA}b). At $S_q \approx 10\%$, however, the scaling decelerates, with solution qualities above $70\%$ for all investigated problem instances at $T=12$. This scaling behavior is primarily influenced by the energy range considered during the algorithm's evolution. Since the GQW cannot sufficiently transport amplitude from all states towards the solution state at short evolution times $T$, the algorithm focuses its efforts on a subset of the graph to maximize the accumulation of amplitude into $\ket{z_{opt}}$. As $T$ increases, the GQW accesses a larger number of states, and consequently, $S_q(T)$ seems to scale according to the amount of accessed amplitude (cf.~disscussion in \secref{subsec:qw:relation_qa}). 

\bigbreak

\paragraph*{Quantum walk:}
For simulating the conventional QW, we employ a procedure similar to that of the GQW. Specifically, we determine the optimal hopping rate $\Gamma$ for each problem instance and $T$ separately, using the Nelder-Mead optimizer with $N_{rep} = 100$ repetitions and a maximum of $N_{opt} = 100$ parameter evaluations, aiming to minimize the final energy expectation value $E_{\Psi}$.

Figure~\ref{fig:ec_Sq_T_GQW-QW-QA}b illustrates the evolution of $S_q(T)$ as a function of $T$ for the QW. Across all problem instances, $S_q(T)$ exhibits a damped oscillation pattern, converging to a stationary solution quality (indicated by dashed lines) below $1$ at $T \geq 12$. Notably, this stationary solution quality decreases exponentially in the problem size $N$, because the QW, with a constant hopping rate $\Gamma$, can only drive a few local subspaces sufficiently (cf.~discussion in \secref{subsec:qw:optimization_problem}). Consequently, the GQW surpasses the QW in terms of performance even for short evolution times (e.g.~$T=0.5$), highlighting the significance of local adjustments to $\Gamma$ already at short time scales. Furthermore, it is noteworthy that the QW is the only algorithm investigated that fails to achieve solution qualities greater than $20\%$ for any problem instance and $T$.

\bigbreak

\paragraph*{Quantum annealing:}
In the context of QA, we examine two annealing schemes: a linear annealing scheme represented by $\Gamma(t) = \left(1 - s(t)\right) / s(t)$, where $s(t) = t/T$, and an optimized locally adiabatic schedule~\cite{LQA_Roland} employing a rescaled time $s^{opt}(t)$. The latter is determined by numerically computing the instantaneous energy gap $\Delta(s)$ between the ground and first excited state of $\hat{H}_{GQW}$ across various values of $s$, followed by solving $ds^{opt}/dt \propto \Delta^2(s^{opt})$ to derive $s^{opt}(t)$. This approach yields an optimized schedule that decelerates the annealing process in regions with small energy gaps while accelerating it elsewhere. A comprehensive discussion on this approach is given in \cite{LQA_Roland}. It's important to note that the linear scheme represents the baseline performance of QA, where no prior optimization phase is required. Conversely, the optimized locally adiabatic schedule mirrors the theoretical best performance of QA for $T \gg 1$, assuming complete knowledge about the optimization problem.

Figure~\ref{fig:ec_Sq_T_GQW-QW-QA}c presents the solution qualities achieved with linear QA (solid lines) and optimized locally adiabatic QA (dashed lines) as a function of $T$. As expected from the adiabatic theorem, $S_q(T)$ increases with the evolution time $T$ for both approaches, with optimized locally adiabatic QA achieving up to one order of magnitude higher solution qualities at $T=12$ then linear QA. Interestingly, for the $N \in \left\{21, 24\right\}$ qubit problems, linear QA beats the optimized locally adiabatic schedule for $T \leq 4$. This is likely caused by diadiabatic transitions in the context of fast annealing. When comparing QA to both the GQW and the QW, we observe a significantly steeper increase in solution quality for the latter two, underscoring the importance of focused amplitude transfers near the solution state for short time scales. For intermediate and long evolution times, QA surpasses the QW. The GQW, however, outperforms both algorithms across all investigated problem instances and $T$. Interestingly, even in the case of long evolutions (e.g., $T=12$) the optimized locally adiabatic QA fails to match the solution qualities achieved by the GQW, indicating the existence of optimal schedules beyond the adiabatic theorem.

\subsection{Performance on large problem instances}
\label{subsec:results:scaling_N}

    \begin{figure}
      \centering
      \includegraphics[width=\columnwidth]{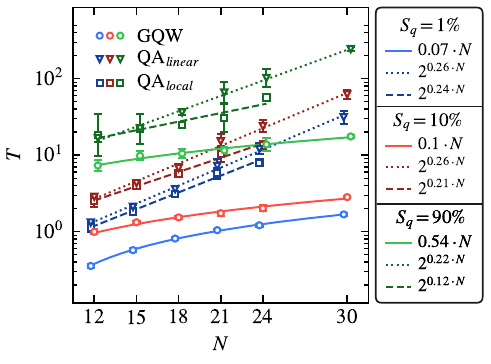}
      \caption{Scaling of the total evolution time $T_{S_q}(N)$ required to achieve a specified solution quality $S_q$ on EC problems as a function of the problem size $N$ for the GQW (circles and solid lines), linear QA (triangles and dotted lines) and optimized locally adiabatic QA~\cite{LQA_Roland} (squares and dashed lines). Colors indicate results for various target solution qualities: $S_q = 1\%$ (blue), $S_q = 10\%$ (red), and $S_q = 90\%$ (green). The solid lines depict linear fits ($a + b \cdot N$), while the dashed and dotted lines represent exponential fits ($a\cdot 2^{b \cdot N}$) applied to the data points (see legend).}
      \label{fig:ec_T_N_GQW-QA}
    \end{figure}

The previous section has demonstrated the efficiency of the proposed guiding procedure for quantum walks in solving combinatorial optimization problems. The GQW outperforms both the QW and QA by providing significantly higher solution qualities across all studied problem instances and evolution times $T$. However, to determine how these algorithms compare for real-world problem sizes ($N \gg 1$) that exceed the capabilities of our numerical simulations, we analyze the scaling of the evolution time $T_{S_q}(N)$ required to reach a solution quality $S_q \in \left\{1\%, 10\%, 90\%\right\}$ as a function of the problem size $N$. The corresponding data is shown in \figref{fig:ec_T_N_GQW-QA}a together with linear ($a + b \cdot N$) and exponential ($a\cdot 2^{b \cdot N}$) fits to the data points obtained by the GQW and QA, respectively. The QW was excluded from this analysis, as it could not reach the required solution qualities. It's worth noting that the QW is not designed as a single-shot algorithm, and we will discuss the multi-shot QW in the subsequent section.

The data demonstrates that quantum annealing (QA) exhibits exponential scaling for both linear and optimized locally adiabatic schedules, with scaling coefficients $b \in \left[0.12, 0.26\right]$ for the three solution quality levels $S_q$. This is in line with the expectation derived from the adiabatic theorem, stating the instantaneous energy gaps, which shrink exponentially in $N$, demand an exponentially slow annealing process for the system to remain in its instantaneous ground state \cite{Young_MEG, Young_FOPT}. Notably, the optimized locally adiabatic QA~\cite{LQA_Roland} achieves the best scaling with $b = 0.12$ for $S_q = 0.9$, showing that local adjustments to the annealing speed can significantly reduce the annealing time needed to reach high solution qualities. In contrast, $T_{S_q}(N)$ follows a linear scaling in $N$ for the GQW. This can be explained by the fact that the depth of a hypercube graph scales linearly in the number of qubits $N$ (i.e., the largest Hamming distance between any two states cannot be larger than $N$). Hence, the GQW must at most drive amplitude transfers within $N$ local subspaces to transport amplitude from any state into the solution state. Since the Hamming distance from $z_{opt}$ to each computational basis state seems to correlate positively with the energy gaps of these states (cf.~\figref{fig:ec_gqw_graph}b), these amplitude transfers are performed simultaneously for all states, yielding a linear in $N$ evolution time $T$. 
    \begin{figure*}
      \centering
      \includegraphics[]{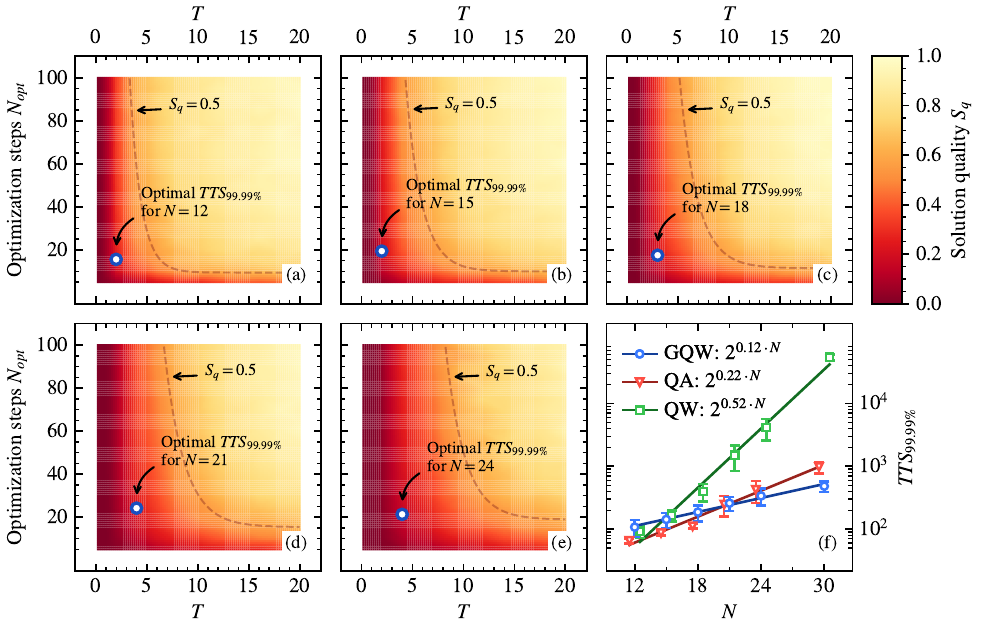}
      \caption{Scaling analysis of the solution quality $S_q$ as a function of the total evolution time $T$ and the number of parameter evaluations $N_{opt}$ performed by the Nelder-Mead algorithm during the classical optimization phase of the hyperparameters $\boldsymbol{\lambda}$ in the GQW. Panels (a)--(e) depict the results for EC problems of sizes $N \in \left\{12, 15, 18, 21, 24\right\}$, respectively. The data is sampled with a granularity of $\Delta T = 1$ and $\Delta N_{opt} = 5$. Each data point represents the average solution quality obtained from $100$ repetitions of the optimization phase. Furthermore, the data is averaged across $10$ random problem instances for each problem size $N$. The blue circles mark the configurations of $T$ and $N_{opt}$ that yield the lowest $TTS_{99.99\%}$ (see \equref{eq:TTS}). The dashed curves indicate exponential fits to the $S_q = 0.5$ contour. In panel (f) we show the scaling of the optimal (smallest) $TTS_{99.99\%}$ achieved by the GQW (blue), linear QA (red), and the QW (green) as a function of the problem size $N$. The solid lines correspond to exponential fits ($a\cdot 2^{b \cdot N}$) applied to the data points.}
      \label{fig:ec_Eval_Sq}
    \end{figure*}
In the regime of large $T$, the GQW can thus be seen as an optimized annealing schedule, where the algorithm selectively spends more time in critical parts of the graph, while progressing faster elsewhere.

Although more extensive studies are required to verify the observation of a linear scaling, our findings identify the GQW as a highly efficient algorithm that achieves high solution qualities within short time scales. This makes the GQW an attractive choice for near-term quantum devices with limited coherence times.

\subsection{Classical optimization phase}
\label{subsec:results:optimization_complexity}
We have examined the performance of the GQW, showcasing its capability to achieve optimal quantum evolutions by guiding amplitude transfers locally in the hypercube graph. Our findings indicate a linear scaling of the total evolution time $T_{S_q}(N)$ with respect to the problem size $N$, surpassing the performance of both the QW and the QA when provided with an optimal set of hyperparameters $\boldsymbol{\lambda}$. However, our investigation has primarily focused on the quantum aspect of this hybrid algorithm, without counting the classical optimization phase responsible for fine-tuning the six hyperparameters $\boldsymbol{\lambda}$.

In Fig.~\ref{fig:ec_Eval_Sq}, we investigate the influence of the classical optimization phase on the GQW by analyzing the scaling of the average solution quality $S_q$ as a function of the total evolution time $T$ and the number of parameter evaluations $N_{opt}$. The latter refers to the number of iterations performed by the Nelder-Mead algorithm during the initial optimization phase. Panels (a)--(e) present the averaged results for the EC problems with sizes $N \in \left\{12, 15, 18, 21, 24\right\}$, respectively.

The data reveals consistent characteristics across all investigated problem instances. Specifically, we observe that the number of parameter evaluations $N_{opt}^{S_q}$ necessary to reach a minimum solution quality $S_q$ decreases exponentially as a function of the total evolution time $T$ (see dashed curves in panels (a)--(e)). Additionally, for a fixed value of $T$, $N_{opt}^{S_q}$ scales exponentially in the problem size $N$. These observations indicate that for short evolutions, where the GQW prepares QW-like hopping rate schedules, the parameter search space tends to be more complex, thereby requiring longer optimization phases. This complexity arises, as the GQW is considering only a few subspaces, making it crucial to precisely set the hyperparameters $\boldsymbol{\lambda}$ to effectively drive amplitude transfers within these subspaces. Notably, the performance of the GQW is particularly sensitive to the choice of $\lambda_4$ and $\lambda_5$ in this regime of $T$. On the other hand, in the case of long evolutions, high-quality solutions can be obtained with just a few iterations from the classical optimizer. This is because deviations from the optimal schedule have minimal impact on the overall evolution of the quantum system, as the optimal hopping rate schedules approach a QA-like evolution, and success is increasingly guaranteed by the adiabatic theorem.

To incorporate the classical optimization phase into our performance evaluations, we consider the time-to-solution
    \begin{align}
        TTS_{P_{target}} = \frac{\ln\left(1 - P_{target}\right)}{\ln\left(1 - P_{gs}\right)} \cdot T + N_{opt} \cdot T,
        \label{eq:TTS}
	\end{align}
where $P_{gs}$ and $T$ denote the success probability and the evolution time of the algorithm, respectively. $TTS_{P_{target}}$ represents the total run time required to measure the solution state $\ket{z_{opt}}$ at least once, with a probability of $P_{target}$, over multiple runs of the algorithm. Note that the optimization phase is accounted for through the offset $N_{opt} \cdot T$.

In Fig.~\ref{fig:ec_Eval_Sq}f, we show a comparison of the scaling of the optimal (smallest) $TTS_{99.99\%}$ achieved by the GQW, the QW, and linear QA as a function of the problem size $N$. Note that optimized locally adiabatic QA is excluded from this analysis, as it requires knowledge of the full spectral information about the optimization problem. Exponential fits ($a \cdot 2^{b \cdot N}$) are included as a reference. The data reveals that for small problem sizes, both the QW and linear QA demonstrate faster convergence to the solution state compared to the GQW, due to the GQW's initial optimization phase. However, for $N \geq 15$ and $N \geq 21$, respectively, the GQW surpasses both algorithms with a scaling factor of $b=0.12$, which is approximately four (two) times better than $b = 0.47$ ($b = 0.22$) for the QW (linear QA). Although the initial optimization phase in the GQW leads to an exponential scaling of $TTS_{99.99\%}$, the GQW's ability to focus solely on a subset of the graph for intermediate values of $T$ enables a significantly more efficient utilization of computational resources compared to the other algorithms. Furthermore, the fact that the exponential scaling is shifted into the optimization phase, while the single run times $T$ scale at most linearly with $N$ (see \secref{subsec:results:scaling_N}), offers the opportunity to distribute the optimization phase across multiple quantum computing devices, thereby providing an option to parallelize the process. This is a feature generally not feasible for QA due to the exponential scaling of $T$, but it could potentially allow solving large optimization problems on near-term quantum devices.


\section{Conclusion}
\label{sec:conclusion}
We have utilized the theory of local amplitude transfer (LAT), which offers a new perspective on the operational principles of quantum annealing (QA) and quantum walks (QWs) beyond the adiabatic theorem, while also providing insights into the design of optimal quantum evolutions. The theory is rooted in the description of a quantum evolution within the eigenspace of the problem Hamiltonian $\hat{H}_C$. In this context, the search space is represented as a graph $G$ where the states are interconnected through the driving Hamiltonian $\hat{H}_D$ (see~\figref{fig:ec_gqw_graph}a). By decomposing $G$ into two-dimensional subspaces spanned by pairs of eigenstates, we have demonstrated that probability amplitude traverses the graph through a sequence of local Rabi oscillations occurring within these subspaces. The amplitude of these oscillations depends on the relative strengths of the local driving and problem Hamiltonians, controlled by the hopping rate $\Gamma$.

We have highlighted that for sufficiently complex problems, the average energy gap of the local problem Hamiltonian monotonically increases as a function of the energy level (see \figref{fig:ec_gqw_graph}b), allowing to selectively drive amplitude transfers within distinct regions of the graph. This property provides a new understanding of how probability amplitude propagates through the search space during continuous-time quantum algorithms.

In particular, we have identified QWs and QA as two formulations of the same underlying principle (see~\figref{fig:ec_evo_T}), with QA corresponding to a sequence of distinct QWs with gradually decreasing hopping rates $\Gamma$. We have shown that a QW-like approach employing a small and constant hopping rate is generally optimal for short evolutions, as it allows for localized dynamics near the solution state. However, it becomes suboptimal for long evolutions, as it fails to effectively transfer amplitude from higher energy states. Conversely, a QA-like strategy is preferred for long evolutions, as it guides amplitude throughout the entire graph, but it is suboptimal for short evolutions, since it spends insufficient time in subspaces near the solution state. Based on these insights, we have argued that optimal quantum evolutions must adapt to the total evolution time $T$ by striking a balance between the number of guided local subspaces (representing the amount of guided amplitude) and the time spent at each energy level (reflecting the amount of amplitude transferred within each local subspace).

Within the LAT framework, we have introduced the guided quantum walk (GQW) as a promising approach for solving large-scale combinatorial optimization problems in the transition region between QWs and QA. The GQW progressively drives local subspaces at gradually decreasing energy levels by utilizing a monotonically decreasing hopping rate $\Gamma(t)$. The hopping rate is controlled through a cubic Bézier curve (see \figref{fig:EC_Hyperparam_Gamma}) defined by six hyperparameters, which allows for fine-tuning the quantum evolution to each problem instance (i.e., energy spectrum of $\hat{H}_C$) and evolution time $T$.

We assessed the performance of the GQW in comparison to QA and QWs on exact cover (EC), traveling salesperson (TSP), and garden optimization (GO) problems ranging from $9$ to $30$ qubits. Across all investigated problem instances and evolution times $T$, the GQW outperformed both the QW and QA significantly. Specifically, at intermediate timescales, our data reveals an up to four (three) orders of magnitude better performance on $30$ qubit problems compared to QA (QW), see~\figref{fig:ec_Sq_T_GQW-QW-QA}. This observation is further supported by the scaling of the minimal evolution time necessary to reach a fixed solution quality as a function of the problem size. In contrast to the exponential scaling observed for QA, the GQW demonstrates a linear scaling, strongly indicating the existence of optimal quantum evolutions that solve combinatorial optimization problems in linear time $T$, thus surpassing the limitations of adiabatic time evolutions.

It is worth noting that the achieved linear scaling is made possible by shifting the exponential scaling to the classical optimization phase of the hyperparameters. Nonetheless, even when considering the parameter tuning in the total run time, the GQW exhibits a time-to-solution scaling that is approximately two (four) times better than for QA (QWs), see~\figref{fig:ec_Eval_Sq}f. This positions the GQW as a powerful tool for deriving optimal annealing schedules. Furthermore, the presence of the exponential scaling in the classical optimization phase, rather than in the single run times, offers the opportunity to distribute the optimization phase across multiple quantum computing devices, thereby enabling parallelization of the process. Moreover, short evolution times also suggest the possibility of discretizing the GQW into a few time steps, therefore adapting the Bézier curve parametrization into a QAOA-like scheme on gate-based quantum computers. These are features generally not feasible for QA due to the exponential scaling of $T$, but it could potentially allow solving large optimization problems on near-term quantum devices. While further investigation is needed to determine how these observations translate to real quantum devices in the presence of environmental noise and on problems with non-canonical energy spectra (e.g., with large degeneracies), our results strongly support the practicality of the GQW for real-world optimization problems, and we expect that our strategy is easily applicable to other types of optimization problems, beyond EC, TSP and GO instances.


\begin{acknowledgments}
The authors thank Viv Kendon, Madita Willsch, Carlos Gonzalez, Berat Yenilen, and Fengping Jin for stimulating discussions.
D.W. acknowledges support from the project J\"ulich UNified Infrastructure for Quantum computing (JUNIQ) that has received funding from the German Federal Ministry of Education and Research (BMBF) and the Ministry of Culture and Science of the State of North Rhine-Westphalia.
The authors gratefully acknowledge the Gauss Centre for Supercomputing e.V. (www.gauss-centre.eu) for funding this project by providing computing time on the GCS Supercomputer JUWELS~\cite{JuwelsClusterBooster} at J\"ulich Supercomputing Centre (JSC).
\end{acknowledgments}

\begin{widetext}
\clearpage
\end{widetext}


\appendix

\section{Problem cost functions}
\label{app:cost_functions}
In this appendix, we provide an overview of the cost functions $C(z)$ for the  EC, TSP, and GO problems under investigation. 

\subsection{Exact cover problem}
\label{subapp:cost_functions:ec}
The EC problem~\cite{Karp1972KarpsNPCompleteProblems,SC_Willsch} involves a set $U = \left\{x_0, x_1, \dots, x_{P-1}\right\}$ with $P$ distinct elements and $N$ subsets $V_i \subseteq U$, such that $U = \bigcup_i V_i$. An EC is a subset $L$ of the set of sets $\left\{V_i\right\}$, such that the elements of $L$ are disjoint sets, and their union is $U$. The problem can be expressed in matrix form using a Boolean problem matrix $A \in \left\{0, 1\right\}^{N \times P}$, where the matrix columns refer to the $P$ elements $x_k$ in $U$ and the matrix rows correspond to the $N$ subsets $V_i$. If $A_{ik} = 1$, then element $x_k$ is included in subset $V_i$, otherwise not. Thus, $L$ is the subset of matrix rows such that the entry $1$ appears in each column exactly once, leading to the cost function
    \begin{align}
        C\left(z\right) &= \sum_{k = 0}^{P - 1} \left[\sum_{i = 0}^{N - 1} A_{ik} z_i - 1\right]^2,
	\end{align}
where $z \in \left\{0, 1\right\}^{\otimes N}$ encodes the selection of subsets $V_i$. Note that the minimum of $C\left(z\right)$ is zero and represents the EC solution. By expanding the square and collecting linear, quadratic, and constant terms in $z_i$, we obtain the QUBO coefficients
    \begin{align}
        Q_{ij} = P +
        \begin{cases}
            \sum_{k = 0}^{P} 2 A_{ik} A_{jk},\quad& i < j\\
            \sum_{k = 0}^{P} A_{ik} \cdot \left(A_{ik} - 2\right) ,\quad& i = j
        \end{cases}.
	\end{align}
	
\subsection{Traveling salesperson problem}
\label{subapp:cost_functions:tsp}
The TSP problem~\cite{Dantzig1954SolutionOfALargeScaleTSP,Bellman1962TSPDynamicProgramming,HeldKarp1962TSP,OperationsResearch,Lucas2014IsingQUBOFormulationManyNPproblems} involves a list of $M$ locations and a cost matrix $c_{kj}$ between each pair of locations $\left(k, j\right)$. The goal is to find the most optimal route, with the lowest total cost, that visits each location exactly once and returns to its starting location. We represent the problem using a Boolean problem matrix $A \in \left\{0, 1\right\}^{M \times M}$, where the matrix rows correspond to the $M$ locations, and the matrix columns denote the order in which the locations are visited. If $A_{k,j} = 1$, then the location $k$ is visited at time step $j$, otherwise not. The solution to the TSP problem is the lowest-cost arrangement of $0$'s and $1$'s in $A$, such that an $1$ is contained in each row and in each column exactly once, leading to the cost function
    \begin{align}
        C\left(z\right) =& \lambda \sum_{k = 0}^{M - 1} \sum_{j = 0}^{M - 1} \sum_{t = 0}^{M - 1} c_{kj} z_{kM + t} z_{jM + t+1}\nonumber\\& + \sum_{k = 0}^{M - 1} \left[\sum_{t = 0}^{M - 1} z_{kM + t} - 1\right]^2\nonumber\\& + \sum_{t = 0}^{M - 1} \left[\sum_{k = 0}^{M - 1} z_{kM + t} - 1\right]^2,
        \label{eq:app:C_TSP}
	\end{align}
with $z_{i} = A_{k,j}$ with $i = kM + j$ and $\lambda$ denoting a free parameter to scale the cost matrix with respect to the constraints. We can determine the QUBO coefficients $Q_{ii'}$ by stepping through the terms in \equref{eq:app:C_TSP} and summing up the respective contributions. It is important to note that we fix the starting location, reducing the number of problem variables $z_i$ to $\left(M - 1\right)^2$. Moreover, the number of optimal routes is even because each route can be traveled in one direction or the other.
	
\subsection{Garden optimization problem}
\label{subapp:cost_functions:go}
The GO problem was first introduced by Gonzalez Calaza et al. \cite{calaza2021gardenoptimization} and refers to the problem of arranging plants in a garden. The garden is represented by $M$ pots randomly distributed on a grid, with a Boolean matrix $J \in \left\{0, 1\right\}^{M \times M}$ encoding the adjacency of pots. Given $P$ distinct plant species, with $c_j$ plants per species, the goal of the GO problem is to find the optimal arrangement of plants in the pots with respect to the relationships between different plant species. These relationships are encoded in the companions matrix $A \in \left\{-1, 0, 1\right\}^{P \times P}$, yielding the cost function
    \begin{align}
        C\left(z\right) =& \sum_{k, k' = 0}^{M - 1} J_{k, k'} \left[1 + \sum_{j, j' = 0}^{P - 1} z_{kP + j} A_{j,j'} z_{k'P + j'}\right]\nonumber\\& + \lambda_1 \sum_{k = 0}^{M - 1} \left[1 - \sum_{j = 0}^{P - 1} z_{kP + j}\right]^2\nonumber\\& + \lambda_2 \sum_{j = 0}^{P - 1} \left[c_j - \sum_{k = 0}^{M - 1} z_{kP + j}\right]^2,
        \label{eq:app:C_GO}
    \end{align}
where $\lambda_1$ and $\lambda_2$ are free parameters. The $M\times P$ Boolean problem variables $z$ indicate whether a plant of species $j$ is placed in pot $k$ (i.e.,~$z_{kP + j}=1$) or not. By stepping through the terms in \equref{eq:app:C_GO} and summing up the respective contributions, we can determine the QUBO coefficients $Q_{ii'}$ in the same way as for the other problems.

\clearpage
\begin{widetext}

\section{LAT theory on general states}
\label{app:gqw_calculations}
In this appendix, we explore the generalized scenario of the LAT theory, where the computational basis states may not be in an equal superposition state. For an in-depth discussion, we refer the reader to \cite{GQW_Master_Thesis}.

Considering the GQW Hamiltonian $\hat{H}_{GQW} = \Gamma(t) \cdot \hat{H}_D + \hat{H}_C$, with the time-dependent hopping rate $\Gamma(t)$, the driving Hamiltonian $\hat{H}_D$ (see \equref{eq:H_D_Hypercube}) and the cost Hamiltonian $\hat{H}_C$ (see \equref{eq:H_C}), the time evolution operator $\hat{U}$ reads

    \begin{align}
        \hat{U}(T) &= \exp \left[ -i \int_{0}^{T} \hat{H}_{GQW}(t) \, dt \right]\\
        &\approx \prod_{k=1}^{P} \; \underbrace{\exp \left[ -i \, \Gamma\left(k \cdot \Delta t\right) \, \Delta t 
        \, \hat{H}_D \right]}_{\hat{U}_D\left(\Gamma\left(k \cdot \Delta t\right) \, \Delta t\right)} \; \underbrace{\exp \left[ -i \, \hat{H}_C \, \Delta t \right]}_{\hat{U}_C\left(\Delta t\right)}.
    \end{align}
Here we have used the first-order Suzuki-Trotter product formula~\cite{SUZUKI1993,Raedt2006} to decompose $\hat{U}$ into small-time steps $\Delta t$. Applying one time-step to a quantum state $\ket{\Psi} = \sum_j \psi_j \ket{j}$ expressed in the computational basis $\left\{\ket{j}\right\}$, yields
    \begin{align}
        \psi_j &= r_j \, e^{-i \alpha_j}\\
        &\xrightarrow[]{\hat{U}_C\left(\gamma\right)} r_j \, e^{-i \, \left(\alpha_j \, + \, \Delta t \, E_j\right)}\\
        &\xrightarrow[]{\hat{U}_D\left(\beta\right)} \sum_{l = 0}^{N} i^l \cdot \cos^{N-l} \left(\beta\right) \sin^{l} \left(\beta\right) \cdot \left[\sum_{\forall k \, : \, \Delta\left(k, j\right) \,  = \, l} r_k \, e^{-i \, \left(\alpha_k \, + \, \Delta t \, E_k\right)}\right]\\
        &\xrightarrow[]{\beta \ll 1} r_j \, e^{-i \, \left(\alpha_j \, + \, \Delta t E_j\right)} \cdot \left[ \cos^{N} \left(\beta\right) + i \cos^{N-1} \left(\beta\right) \sin \left(\beta\right) \sum_{\forall k \, : \, \Delta\left(k, j\right) \, = \, 1} \frac{r_k}{r_j} \, e^{-i \, \left( \Delta\alpha_{k, j} \, + \, \Delta t \, \Delta E_{k, j}\right)}\right],
    \end{align}
with $\beta = \Gamma\left(k \, \Delta t\right) \cdot \Delta t$, $\gamma = \Delta t$, $\Delta\alpha_{k, j} = \alpha_k - \alpha_j$, $\Delta E_{k, j} = E_k - E_j = \bra{k}\hat{H}_C\ket{k} - \bra{j}\hat{H}_C\ket{j}$, and $\Delta\left(k, j\right)$ denoting the Hamming distance between the labels $k$ and $j$. In the limit of $\Delta t \ll 1$, $\hat{H}_D$ simplifies to first-order interactions, enabling amplitude transfers solely between states of Hamming distance one and thus revealing the hypercube graph (e.g. see \figref{fig:ec_gqw_graph}a). Considering one of these two-dimensional subspaces spanned by the basis states $\ket{j}$ and $\ket{k}$ (i.e., $\Delta\left(k, j\right) \, = \, 1$), the amplitude transfer during one time-step is given by
    \begin{align}
        \psi_j &= \; r_j \; e^{-i \; \alpha_j}\\
        &\xrightarrow[]{\hat{U}\left(\Delta t\right)} r_j \; e^{-i \; \left(\alpha_j \; + \; \gamma \cdot E_j \right)} \Bigl[\cos^N \left(\beta\right)\notag\\
        &\quad\quad\quad\quad\quad\quad\quad\quad\quad\quad\quad \; +\frac{r_k}{r_j}\cos^{N-1}\left(\beta\right) \sin\left(\beta\right) \sin\left(\Delta \alpha_{k, j} + \gamma \cdot \Delta E_{k, j}\right)\notag\\
        &\quad\quad\quad\quad\quad\quad\quad\quad\quad\quad\quad \; + i \; \frac{r_k}{r_j}\cos^{N-1}\left(\beta\right) \sin\left(\beta\right) \cos\left(\Delta \alpha_{k, j} + \gamma \cdot \Delta E_{k, j}\right)\Bigr],
        \label{eq:app:lat_j}\\
        \psi_k &= \; r_k \; e^{-i \; \alpha_k}\\
        &\xrightarrow[]{\hat{U}\left(\Delta t\right)} r_k \; e^{-i \; \left(\alpha_k \; + \; \gamma \cdot E_k \right)} \Bigl[\cos^N \left(\beta\right)\notag\\
        &\quad\quad\quad\quad\quad\quad\quad\quad\quad\quad\quad \; -\frac{r_j}{r_k}\cos^{N-1}\left(\beta\right) \sin\left(\beta\right) \sin\left(\Delta \alpha_{k, j} + \gamma \cdot \Delta E_{k, j}\right)\notag\\
        &\quad\quad\quad\quad\quad\quad\quad\quad\quad\quad\quad \; + i \; \frac{r_j}{r_k}\cos^{N-1}\left(\beta\right) \sin\left(\beta\right) \cos\left(\Delta \alpha_{k, j} + \gamma \cdot \Delta E_{k, j}\right)\Bigr].
        \label{eq:app:lat_k}
    \end{align}

Equations \ref{eq:app:lat_j} and \ref{eq:app:lat_k} demonstrate that complex phase shifts caused by $\hat{H}_C$ result in opposing signs of the second terms of the squared brackets, with the lower (higher) energy state having a '$+$' ('$-$'). When $r_j/r_k \approx 1$ and $\beta$ is small, the amplitude of each basis state after one time-step is mainly defined by $\cos^N \left(\beta\right)$, which denotes the amount of amplitude that remains at each basis state. The second term increases this amplitude for the low energy state, while decreasing it for the high energy state, resulting in a localized amplitude transfer between the two. As a result, the hypercube graph becomes directed and probability amplitude is steered towards the ground state of $\hat{H}_C$. It is important to note, however, that for a time-independent $\beta$ (e.g. in a QW) this process will eventually reverse, causing amplitude to flow back into the high energy state. This occurs because the fraction $r_j / r_k$ grows as amplitude is exchanged between the two states. Thus, the higher energy state's second term eventually overwhelms the $\cos^N \left(\beta\right)$ resulting in an increased amplitude, while the lower energy state's second term becomes suppressed causing an overall decrease in amplitude as $\cos^N \left(\beta\right) < 1$. This leads to oscillations in the probability amplitude between the two states (see e.g. \figref{fig:qw_exact_cover_problem}). To prevent this, the GQW makes use of a time-dependent monotonically decreasing hopping rate $\Gamma(t)$, which counteracts the increasing fraction $r_j / r_k$ through the $\sin \left(\beta\right)$. By doing so, the GQW is able to prohibit the back-propagation of amplitude even in the case of large detuning by suppressing the amplitude transfer locally.

\color{black}
\clearpage
\end{widetext}

\section{GO and TSP results}
\label{app:go_tsp_results}
In this appendix, we present the simulation results for the GQW, the QW, and QA applied to GO and TSP problems. We investigate GO problems with problem sizes $N \in \left\{12, 15, 18, 21, 24, 30\right\}$ and TSP problems with problem sizes $N \in \left\{9, 16, 25\right\}$, considering a total of $10$ randomly generated instances for each problem size and type. Similar to the EC results from the main text, the GQW and the QW are tuned for each problem instance and evolution time $T$ separately by optimizing their hyperparameters to minimize the energy expectation value $E_{\Psi}$ of the final quantum state $\ket{\Psi}$. Each set of hyperparameters is selected from a pool of $N_{rep} = 100$ runs of the Nelder-Mead classical optimizer, where for each run, the hyperparameters are initialized randomly and adjusted up to a maximum of $N_{eval} = 100$ times. The QA results are obtained using a linear annealing scheme.

Figures~\ref{fig:go_Sq_T_GQW-QW-QA} and \ref{fig:tsp_Sq_T_GQW-QW-QA} depict the scaling of the solution quality $S_q$ as a function of the evolution time $T$ and the problem size $N$ for the three algorithms on GO and TSP problems, respectively. The data exhibit similar characteristics to the EC problems presented in \figref{fig:ec_Sq_T_GQW-QW-QA} and are qualitatively consistent with the discussion in \secref{subsec:results:comparison}. Notably, the GQW achieves higher solution qualities for some large GO problems (e.g., $N=30$) compared to some smaller problem sizes (e.g., $N=18$) for long evolutions (see inset in \figref{fig:go_Sq_T_GQW-QW-QA}a). This indicates that the hardness of the combinatorial optimization problems varies throughout the problem sizes, such that some large problem instances are easier to solve than their small counterparts once the GQW considers the entire graph. Consequently, we expect similar characteristics in the scaling of $S_q$ to appear for QA for long evolutions.

In Figures~\ref{fig:go_T_N_GQW-QA} and \ref{fig:tsp_T_N_GQW-QA}, we present the scaling of the minimum evolution time $T_{S_q}(N)$ required to achieve a specified solution quality $S_q$ as a function of the problem size $N$. Similar to the results obtained for the EC problems, $T_{S_q}(N)$ exhibits linear scaling for the GQW, in contrast to the exponential scaling observed for QA. This further supports the existence of optimal hopping rate schedules that can solve combinatorial optimization problems in linear time. Note, however, that for the GO problems, $T_{0.9}(N)$ neither corresponds to an exponential nor a linear scaling, due to the differences in the hardness of the optimization problems.

Figures~\ref{fig:go_TTS} and \ref{fig:tsp_TTS} show the scaling of the time-to-solution (see \equref{eq:TTS}) as a function of the problem size $N$. For both problem types, the GQW achieves a superior scaling compared to QA and the QW by factors of $\approx 2$ (QA on GO problems), $\approx 4$ (QW on GO problems) and $\approx 2$ (QA and QW on TSP problems). 

\begin{widetext}
\clearpage

    \begin{figure*}
        \centering
        \includegraphics[]{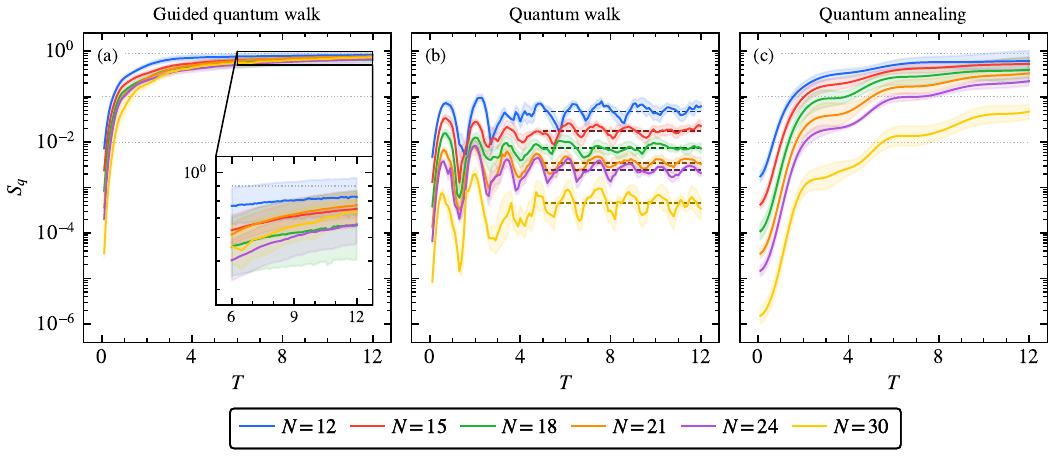}
        \caption{Performance comparison of (a) the GQW, (b) the QW, and (c) QA as a function of the total evolution time $T$ on GO problems with problem sizes $N \in \left\{12, 15, 18, 21, 24, 30\right\}$ (see legend). The performance is assessed based on the solution quality $S_q$ defined in \eqref{eq:S}. Each problem size $N$ is evaluated using a set of $10$ randomly generated instances, and the presented data corresponds to the geometric mean solution qualities (solid curves) along with the geometric standard deviations (shaded areas). Note that the GQW and the QW undergo a prior classical optimization phase. The dotted gray lines indicate horizontal cuts ($S_q \in \left\{1\%, 10\%, 90\%\right\}$), which are detailed in \figref{fig:go_T_N_GQW-QA}. The inset in panel (a) provides a zoomed-in view of the region corresponding to large evolution times $T$. Additionally, the dashed lines in panel (b) indicate the stationary solution quality that the damped oscillation of $S_q(T)$ approaches for the QW at large $T$.}
        \label{fig:go_Sq_T_GQW-QW-QA}
    \end{figure*}

    \begin{figure}[p]
      \centering
      \includegraphics[width=0.5\textwidth]{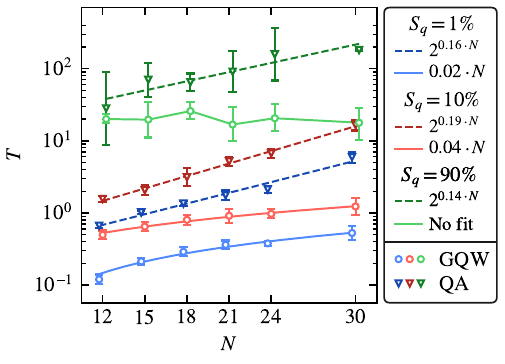}
      \caption{Scaling behavior of the total evolution time $T_{S_q}(N)$ required to achieve a specified solution quality $S_q$ on GO problems as a function of the problem size $N$ for the GQW (circles) and linear QA (triangles). Colors indicate results for various target solution qualities: $S_q = 1\%$ (blue), $S_q = 10\%$ (red), and $S_q = 90\%$ (green). The solid lines depict linear fits ($a + b \cdot N$), while the dashed lines represent exponential fits ($a\cdot 2^{b \cdot N}$) applied to the data points (see legend).}
      \label{fig:go_T_N_GQW-QA}
    \end{figure}\qquad%
    \begin{figure}[p]
      \centering
     \includegraphics[width=0.5\textwidth]{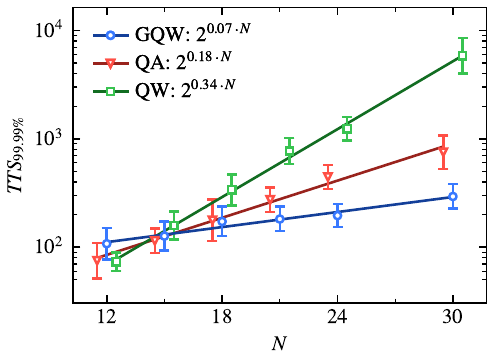}
      \caption{Scaling analysis of the optimal (smallest) time-to-solution $TTS_{99.99\%}$ (see \equref{eq:TTS}) for the GQW (blue), QA (red), and the QW (green) on GO problems as a function of the problem size $N$. The solid lines correspond to exponential fits ($a\cdot 2^{b \cdot N}$) applied to the data points (see legend).}
      \label{fig:go_TTS}
    \end{figure}
    
    \clearpage

    \begin{figure*}
        \centering
        \includegraphics[]{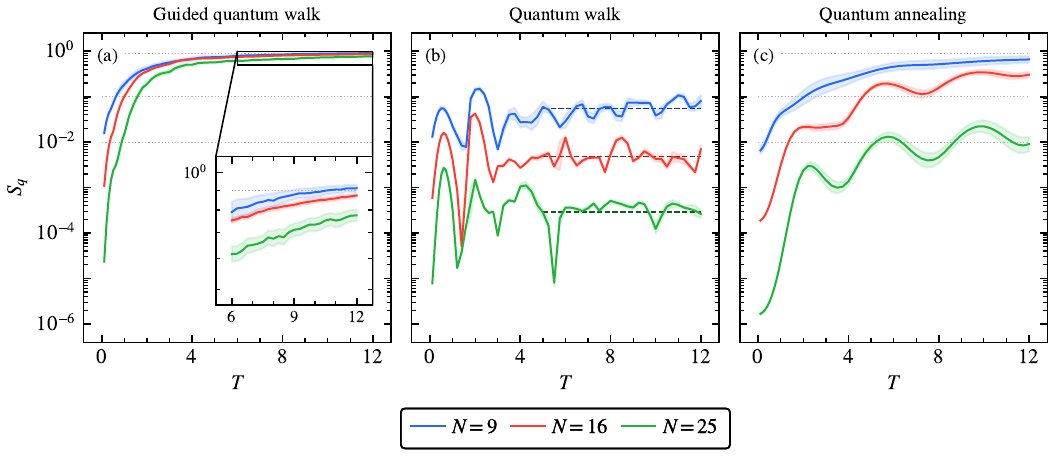}
        \caption{Performance comparison of (a) the GQW, (b) the QW, and (c) QA as a function of the total evolution time $T$ on TSP problems with problem sizes $N \in \left\{9, 16, 25\right\}$ (see legend). The performance is assessed based on the solution quality $S_q$ defined in \eqref{eq:S}. Each problem size $N$ is evaluated using a set of $10$ randomly generated instances, and the presented data corresponds to the geometric mean solution qualities (solid curves) along with the geometric standard deviations (shaded areas). Note that the GQW and the QW undergo a prior classical optimization phase. The dotted gray lines indicate horizontal cuts ($S_q \in \left\{1\%, 10\%, 90\%\right\}$), which are detailed in \figref{fig:tsp_T_N_GQW-QA}. The inset in panel (a) provides a zoomed-in view of the region corresponding to large evolution times $T$. Additionally, the dashed lines in panel (b) indicate the stationary solution quality that the damped oscillation of $S_q(T)$ approaches for the QW at large $T$.}
        \label{fig:tsp_Sq_T_GQW-QW-QA}
    \end{figure*}
    
    \begin{figure}[p]
          \centering
          \includegraphics[width=0.5\textwidth]{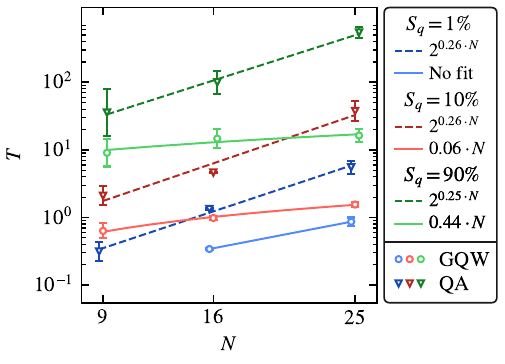}
          \caption{Scaling behavior of the total evolution time $T_{S_q}(N)$ required to achieve a specified solution quality $S_q$ on TSP problems as a function of the problem size $N$ for the GQW (circles) and linear QA (triangles). Colors indicate results for various target solution qualities: $S_q = 1\%$ (blue), $S_q = 10\%$ (red), and $S_q = 90\%$ (green). The solid lines depict linear fits ($a + b \cdot N$), while the dashed lines represent exponential fits ($a\cdot 2^{b \cdot N}$) applied to the data points (see legend).}
          \label{fig:tsp_T_N_GQW-QA}
        \end{figure}\qquad%
        \begin{figure}[p]
          \centering
         \includegraphics[width=0.5\textwidth]{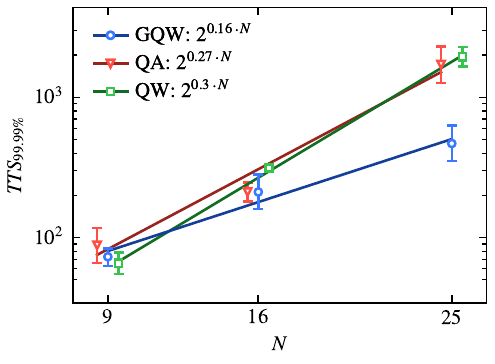}
          \caption{Scaling analysis of the optimal (smallest) time-to-solution $TTS_{99.99\%}$ (see \equref{eq:TTS}) for the GQW (blue), QA (red), and the QW (green) on TSP problems as a function of the problem size $N$. The solid lines correspond to exponential fits ($a\cdot 2^{b \cdot N}$) applied to the data points (see legend).}
          \label{fig:tsp_TTS}
        \end{figure}
    
    \clearpage

\end{widetext}


\bibliographystyle{apsrev4-2custom}
\bibliography{bibliography}

\begin{thebibliography}{55}%
\makeatletter
\providecommand \@ifxundefined [1]{%
 \@ifx{#1\undefined}
}%
\providecommand \@ifnum [1]{%
 \ifnum #1\expandafter \@firstoftwo
 \else \expandafter \@secondoftwo
 \fi
}%
\providecommand \@ifx [1]{%
 \ifx #1\expandafter \@firstoftwo
 \else \expandafter \@secondoftwo
 \fi
}%
\providecommand \natexlab [1]{#1}%
\providecommand \emph  [1]{``#1''}%
\providecommand \bibnamefont  [1]{#1}%
\providecommand \bibfnamefont [1]{#1}%
\providecommand \citenamefont [1]{#1}%
\providecommand \href@noop [0]{\@secondoftwo}%
\providecommand \href [0]{\begingroup \@sanitize@url \@href}%
\providecommand \@href[1]{\@@startlink{#1}\@@href}%
\providecommand \@@href[1]{\endgroup#1\@@endlink}%
\providecommand \@sanitize@url [0]{\catcode `\\12\catcode `\$12\catcode
  `\&12\catcode `\#12\catcode `\^12\catcode `\_12\catcode `\%12\relax}%
\providecommand \@@startlink[1]{}%
\providecommand \@@endlink[0]{}%
\providecommand \url  [0]{\begingroup\@sanitize@url \@url }%
\providecommand \@url [1]{\endgroup\@href {#1}{\urlprefix }}%
\providecommand \urlprefix  [0]{URL }%
\providecommand \Eprint [0]{\href }%
\providecommand \doibase [0]{https://doi.org/}%
\providecommand \selectlanguage [0]{\@gobble}%
\providecommand \bibinfo  [0]{\@secondoftwo}%
\providecommand \bibfield  [0]{\@secondoftwo}%
\providecommand \translation [1]{[#1]}%
\providecommand \BibitemOpen [0]{}%
\providecommand \bibitemStop [0]{}%
\providecommand \bibitemNoStop [0]{.\EOS\space}%
\providecommand \EOS [0]{\spacefactor3000\relax}%
\providecommand \BibitemShut  [1]{\csname bibitem#1\endcsname}%
\let\auto@bib@innerbib\@empty
\bibitem [{\citenamefont {Marzec}(2016)}]{PortfolioOpt_Marzec}%
  \BibitemOpen
  \bibfield  {author} {\bibinfo {author} {\bibfnamefont {M.}~\bibnamefont
  {Marzec}},\ }\bibfield  {title} {\emph {\bibinfo {title} {Portfolio
  Optimization: Applications in Quantum Computing},}\ }\href
  {https://doi.org/10.1002/9781118593486.ch4} {\bibfield  {journal} {\bibinfo
  {journal} {Handbook of High‐Frequency Trading and Modeling in Finance}\ ,\
  \bibinfo {pages} {73}} (\bibinfo {year} {2016})}\BibitemShut {NoStop}%
\bibitem [{\citenamefont {Willsch}\ \emph {et~al.}(2022)\citenamefont
  {Willsch}, \citenamefont {Willsch}, \citenamefont {Calaza}, \citenamefont
  {Jin}, \citenamefont {{De Raedt}}, \citenamefont {Svensson},\ and\
  \citenamefont {Michielsen}}]{SC_Willsch}%
  \BibitemOpen
  \bibfield  {author} {\bibinfo {author} {\bibfnamefont {D.}~\bibnamefont
  {Willsch}}, \bibinfo {author} {\bibfnamefont {M.}~\bibnamefont {Willsch}},
  \bibinfo {author} {\bibfnamefont {C.~D.~G.}\ \bibnamefont {Calaza}}, \bibinfo
  {author} {\bibfnamefont {F.}~\bibnamefont {Jin}}, \bibinfo {author}
  {\bibfnamefont {H.}~\bibnamefont {{De Raedt}}}, \bibinfo {author}
  {\bibfnamefont {M.}~\bibnamefont {Svensson}},\ and\ \bibinfo {author}
  {\bibfnamefont {K.}~\bibnamefont {Michielsen}},\ }\bibfield  {title} {\emph
  {\bibinfo {title} {Benchmarking Advantage and D-Wave 2000Q quantum annealers
  with exact cover problems},}\ }\bibfield  {journal} {\bibinfo  {journal}
  {Quantum Information Processing}\ }\textbf {\bibinfo {volume} {21}},\ \href
  {https://doi.org/10.1007/s11128-022-03476-y} {10.1007/s11128-022-03476-y}
  (\bibinfo {year} {2022})\BibitemShut {NoStop}%
\bibitem [{\citenamefont {Amin}\ \emph {et~al.}(2018)\citenamefont {Amin},
  \citenamefont {Andriyash}, \citenamefont {Rolfe}, \citenamefont
  {Kulchytskyy},\ and\ \citenamefont {Melko}}]{ML_Amin}%
  \BibitemOpen
  \bibfield  {author} {\bibinfo {author} {\bibfnamefont {M.~H.}\ \bibnamefont
  {Amin}}, \bibinfo {author} {\bibfnamefont {E.}~\bibnamefont {Andriyash}},
  \bibinfo {author} {\bibfnamefont {J.}~\bibnamefont {Rolfe}}, \bibinfo
  {author} {\bibfnamefont {B.}~\bibnamefont {Kulchytskyy}},\ and\ \bibinfo
  {author} {\bibfnamefont {R.}~\bibnamefont {Melko}},\ }\bibfield  {title}
  {\emph {\bibinfo {title} {Quantum Boltzmann Machine},}\ }\href
  {https://doi.org/10.1103/physrevx.8.021050} {\bibfield  {journal} {\bibinfo
  {journal} {Phys. Rev. X}\ }\textbf {\bibinfo {volume} {8}},\ \bibinfo {pages}
  {021050} (\bibinfo {year} {2018})}\BibitemShut {NoStop}%
\bibitem [{\citenamefont {Benedetti}\ \emph {et~al.}(2016)\citenamefont
  {Benedetti}, \citenamefont {Realpe-G\'omez}, \citenamefont {Biswas},\ and\
  \citenamefont {Perdomo-Ortiz}}]{ML_Benedetti}%
  \BibitemOpen
  \bibfield  {author} {\bibinfo {author} {\bibfnamefont {M.}~\bibnamefont
  {Benedetti}}, \bibinfo {author} {\bibfnamefont {J.}~\bibnamefont
  {Realpe-G\'omez}}, \bibinfo {author} {\bibfnamefont {R.}~\bibnamefont
  {Biswas}},\ and\ \bibinfo {author} {\bibfnamefont {A.}~\bibnamefont
  {Perdomo-Ortiz}},\ }\bibfield  {title} {\emph {\bibinfo {title} {Estimation
  of effective temperatures in quantum annealers for sampling applications: A
  case study with possible applications in deep learning},}\ }\href
  {https://doi.org/10.1103/PhysRevA.94.022308} {\bibfield  {journal} {\bibinfo
  {journal} {Phys. Rev. A}\ }\textbf {\bibinfo {volume} {94}},\ \bibinfo
  {pages} {022308} (\bibinfo {year} {2016})}\BibitemShut {NoStop}%
\bibitem [{\citenamefont {Childs}\ and\ \citenamefont
  {Goldstone}(2004)}]{Childs2004CTQWSearchProblem}%
  \BibitemOpen
  \bibfield  {author} {\bibinfo {author} {\bibfnamefont {A.~M.}\ \bibnamefont
  {Childs}}\ and\ \bibinfo {author} {\bibfnamefont {J.}~\bibnamefont
  {Goldstone}},\ }\bibfield  {title} {\emph {\bibinfo {title} {Spatial search
  by quantum walk},}\ }\href {https://doi.org/10.1103/PhysRevA.70.022314}
  {\bibfield  {journal} {\bibinfo  {journal} {Phys. Rev. A}\ }\textbf {\bibinfo
  {volume} {70}},\ \bibinfo {pages} {022314} (\bibinfo {year}
  {2004})}\BibitemShut {NoStop}%
\bibitem [{\citenamefont {Perdomo-Ortiz}\ \emph {et~al.}(2012)\citenamefont
  {Perdomo-Ortiz}, \citenamefont {Dickson}, \citenamefont {Drew-Brook},
  \citenamefont {Rose},\ and\ \citenamefont {Aspuru-Guzik}}]{CB_Perdomo}%
  \BibitemOpen
  \bibfield  {author} {\bibinfo {author} {\bibfnamefont {A.}~\bibnamefont
  {Perdomo-Ortiz}}, \bibinfo {author} {\bibfnamefont {N.}~\bibnamefont
  {Dickson}}, \bibinfo {author} {\bibfnamefont {M.}~\bibnamefont {Drew-Brook}},
  \bibinfo {author} {\bibfnamefont {G.}~\bibnamefont {Rose}},\ and\ \bibinfo
  {author} {\bibfnamefont {A.}~\bibnamefont {Aspuru-Guzik}},\ }\bibfield
  {title} {\emph {\bibinfo {title} {Finding low-energy conformations of lattice
  protein models by quantum annealing},}\ }\bibfield  {journal} {\bibinfo
  {journal} {Scientific Reports}\ }\href {https://doi.org/10.1038/srep00571}
  {10.1038/srep00571} (\bibinfo {year} {2012})\BibitemShut {NoStop}%
\bibitem [{\citenamefont {Boucherie}\ \emph {et~al.}(2021)\citenamefont
  {Boucherie}, \citenamefont {Braaksma},\ and\ \citenamefont
  {Tijms}}]{OperationsResearch}%
  \BibitemOpen
  \bibfield  {author} {\bibinfo {author} {\bibfnamefont {R.~J.}\ \bibnamefont
  {Boucherie}}, \bibinfo {author} {\bibfnamefont {A.}~\bibnamefont
  {Braaksma}},\ and\ \bibinfo {author} {\bibfnamefont {H.}~\bibnamefont
  {Tijms}},\ }\href {https://doi.org/10.1142/12343} {\emph {\bibinfo {title}
  {Operations Research}}}\ (\bibinfo  {publisher} {World Scientific},\ \bibinfo
  {year} {2021})\BibitemShut {NoStop}%
\bibitem [{\citenamefont {Preskill}(2018)}]{Preskill2018NISQ}%
  \BibitemOpen
  \bibfield  {author} {\bibinfo {author} {\bibfnamefont {J.}~\bibnamefont
  {Preskill}},\ }\bibfield  {title} {\emph {\bibinfo {title} {Quantum
  {C}omputing in the {NISQ} era and beyond},}\ }\href
  {https://doi.org/10.22331/q-2018-08-06-79} {\bibfield  {journal} {\bibinfo
  {journal} {{Quantum}}\ }\textbf {\bibinfo {volume} {2}},\ \bibinfo {pages}
  {79} (\bibinfo {year} {2018})}\BibitemShut {NoStop}%
\bibitem [{\citenamefont {Aharonov}\ \emph {et~al.}(1993)\citenamefont
  {Aharonov}, \citenamefont {Davidovich},\ and\ \citenamefont
  {Zagury}}]{Aharonov1993QuantumRandomWalks}%
  \BibitemOpen
  \bibfield  {author} {\bibinfo {author} {\bibfnamefont {Y.}~\bibnamefont
  {Aharonov}}, \bibinfo {author} {\bibfnamefont {L.}~\bibnamefont
  {Davidovich}},\ and\ \bibinfo {author} {\bibfnamefont {N.}~\bibnamefont
  {Zagury}},\ }\bibfield  {title} {\emph {\bibinfo {title} {Quantum random
  walks},}\ }\href {https://doi.org/10.1103/PhysRevA.48.1687} {\bibfield
  {journal} {\bibinfo  {journal} {Phys. Rev. A}\ }\textbf {\bibinfo {volume}
  {48}},\ \bibinfo {pages} {1687} (\bibinfo {year} {1993})}\BibitemShut
  {NoStop}%
\bibitem [{\citenamefont {Childs}\ \emph {et~al.}(2002)\citenamefont {Childs},
  \citenamefont {Deotto}, \citenamefont {Farhi}, \citenamefont {Goldstone},
  \citenamefont {Gutmann},\ and\ \citenamefont
  {Landahl}}]{Childs2002QuantumSearchByMeasurement}%
  \BibitemOpen
  \bibfield  {author} {\bibinfo {author} {\bibfnamefont {A.~M.}\ \bibnamefont
  {Childs}}, \bibinfo {author} {\bibfnamefont {E.}~\bibnamefont {Deotto}},
  \bibinfo {author} {\bibfnamefont {E.}~\bibnamefont {Farhi}}, \bibinfo
  {author} {\bibfnamefont {J.}~\bibnamefont {Goldstone}}, \bibinfo {author}
  {\bibfnamefont {S.}~\bibnamefont {Gutmann}},\ and\ \bibinfo {author}
  {\bibfnamefont {A.~J.}\ \bibnamefont {Landahl}},\ }\bibfield  {title} {\emph
  {\bibinfo {title} {Quantum search by measurement},}\ }\href
  {https://doi.org/10.1103/PhysRevA.66.032314} {\bibfield  {journal} {\bibinfo
  {journal} {Phys. Rev. A}\ }\textbf {\bibinfo {volume} {66}},\ \bibinfo
  {pages} {032314} (\bibinfo {year} {2002})}\BibitemShut {NoStop}%
\bibitem [{\citenamefont {Callison}\ \emph {et~al.}(2021)\citenamefont
  {Callison}, \citenamefont {Festenstein}, \citenamefont {Chen}, \citenamefont
  {Nita}, \citenamefont {Kendon},\ and\ \citenamefont
  {Chancellor}}]{Callison2021EnergeticPerspectiveQAMultiStageQW}%
  \BibitemOpen
  \bibfield  {author} {\bibinfo {author} {\bibfnamefont {A.}~\bibnamefont
  {Callison}}, \bibinfo {author} {\bibfnamefont {M.}~\bibnamefont
  {Festenstein}}, \bibinfo {author} {\bibfnamefont {J.}~\bibnamefont {Chen}},
  \bibinfo {author} {\bibfnamefont {L.}~\bibnamefont {Nita}}, \bibinfo {author}
  {\bibfnamefont {V.}~\bibnamefont {Kendon}},\ and\ \bibinfo {author}
  {\bibfnamefont {N.}~\bibnamefont {Chancellor}},\ }\bibfield  {title} {\emph
  {\bibinfo {title} {Energetic Perspective on Rapid Quenches in Quantum
  Annealing},}\ }\href {https://doi.org/10.1103/PRXQuantum.2.010338} {\bibfield
   {journal} {\bibinfo  {journal} {PRX Quantum}\ }\textbf {\bibinfo {volume}
  {2}},\ \bibinfo {pages} {010338} (\bibinfo {year} {2021})}\BibitemShut
  {NoStop}%
\bibitem [{\citenamefont {Banks}\ \emph {et~al.}(2023)\citenamefont {Banks},
  \citenamefont {Haque}, \citenamefont {Nazef}, \citenamefont {Fethallah},
  \citenamefont {Ruqaya}, \citenamefont {Ahsan}, \citenamefont {Vora},
  \citenamefont {Tahir}, \citenamefont {Ahmad}, \citenamefont {Hewins},
  \citenamefont {Shah}, \citenamefont {Baranwal}, \citenamefont {Arora},
  \citenamefont {Asad}, \citenamefont {Khan}, \citenamefont {Hasan},
  \citenamefont {Azad}, \citenamefont {Fedaiee}, \citenamefont {Majeed},
  \citenamefont {Bhuyan}, \citenamefont {Tarannum}, \citenamefont {Ali},
  \citenamefont {Browne},\ and\ \citenamefont
  {Warburton}}]{banks2023CTQWforMAXCUTarehot}%
  \BibitemOpen
  \bibfield  {author} {\bibinfo {author} {\bibfnamefont {R.~J.}\ \bibnamefont
  {Banks}}, \bibinfo {author} {\bibfnamefont {E.}~\bibnamefont {Haque}},
  \bibinfo {author} {\bibfnamefont {F.}~\bibnamefont {Nazef}}, \bibinfo
  {author} {\bibfnamefont {F.}~\bibnamefont {Fethallah}}, \bibinfo {author}
  {\bibfnamefont {F.}~\bibnamefont {Ruqaya}}, \bibinfo {author} {\bibfnamefont
  {H.}~\bibnamefont {Ahsan}}, \bibinfo {author} {\bibfnamefont
  {H.}~\bibnamefont {Vora}}, \bibinfo {author} {\bibfnamefont {H.}~\bibnamefont
  {Tahir}}, \bibinfo {author} {\bibfnamefont {I.}~\bibnamefont {Ahmad}},
  \bibinfo {author} {\bibfnamefont {I.}~\bibnamefont {Hewins}}, \bibinfo
  {author} {\bibfnamefont {I.}~\bibnamefont {Shah}}, \bibinfo {author}
  {\bibfnamefont {K.}~\bibnamefont {Baranwal}}, \bibinfo {author}
  {\bibfnamefont {M.}~\bibnamefont {Arora}}, \bibinfo {author} {\bibfnamefont
  {M.}~\bibnamefont {Asad}}, \bibinfo {author} {\bibfnamefont {M.}~\bibnamefont
  {Khan}}, \bibinfo {author} {\bibfnamefont {N.}~\bibnamefont {Hasan}},
  \bibinfo {author} {\bibfnamefont {N.}~\bibnamefont {Azad}}, \bibinfo {author}
  {\bibfnamefont {S.}~\bibnamefont {Fedaiee}}, \bibinfo {author} {\bibfnamefont
  {S.}~\bibnamefont {Majeed}}, \bibinfo {author} {\bibfnamefont
  {S.}~\bibnamefont {Bhuyan}}, \bibinfo {author} {\bibfnamefont
  {T.}~\bibnamefont {Tarannum}}, \bibinfo {author} {\bibfnamefont
  {Y.}~\bibnamefont {Ali}}, \bibinfo {author} {\bibfnamefont {D.~E.}\
  \bibnamefont {Browne}},\ and\ \bibinfo {author} {\bibfnamefont {P.~A.}\
  \bibnamefont {Warburton}},\ }\bibfield  {title} {\emph {\bibinfo {title}
  {Continuous-time quantum walks for MAX-CUT are hot},}\ }\Eprint
  {https://arxiv.org/abs/2306.10365} {arXiv:2306.10365 [quant-ph]}  (\bibinfo
  {year} {2023})\BibitemShut {NoStop}%
\bibitem [{\citenamefont {Apolloni}\ \emph {et~al.}(1989)\citenamefont
  {Apolloni}, \citenamefont {Carvalho},\ and\ \citenamefont
  {de~Falco}}]{Apolloni89}%
  \BibitemOpen
  \bibfield  {author} {\bibinfo {author} {\bibfnamefont {B.}~\bibnamefont
  {Apolloni}}, \bibinfo {author} {\bibfnamefont {C.}~\bibnamefont {Carvalho}},\
  and\ \bibinfo {author} {\bibfnamefont {D.}~\bibnamefont {de~Falco}},\
  }\bibfield  {title} {\emph {\bibinfo {title} {Quantum stochastic
  optimization},}\ }\href {https://doi.org/10.1016/0304-4149(89)90040-9}
  {\bibfield  {journal} {\bibinfo  {journal} {Stoch. Process. Their Appl.}\
  }\textbf {\bibinfo {volume} {33}},\ \bibinfo {pages} {233 } (\bibinfo {year}
  {1989})}\BibitemShut {NoStop}%
\bibitem [{\citenamefont {Finnila}\ \emph {et~al.}(1994)\citenamefont
  {Finnila}, \citenamefont {Gomez}, \citenamefont {Sebenik}, \citenamefont
  {Stenson},\ and\ \citenamefont {Doll}}]{finnila94}%
  \BibitemOpen
  \bibfield  {author} {\bibinfo {author} {\bibfnamefont {A.}~\bibnamefont
  {Finnila}}, \bibinfo {author} {\bibfnamefont {M.}~\bibnamefont {Gomez}},
  \bibinfo {author} {\bibfnamefont {C.}~\bibnamefont {Sebenik}}, \bibinfo
  {author} {\bibfnamefont {C.}~\bibnamefont {Stenson}},\ and\ \bibinfo {author}
  {\bibfnamefont {J.}~\bibnamefont {Doll}},\ }\bibfield  {title} {\emph
  {\bibinfo {title} {Quantum annealing: A new method for minimizing
  multidimensional functions},}\ }\href
  {https://doi.org/10.1016/0009-2614(94)00117-0} {\bibfield  {journal}
  {\bibinfo  {journal} {Chem. Phys. Lett.}\ }\textbf {\bibinfo {volume}
  {219}},\ \bibinfo {pages} {343} (\bibinfo {year} {1994})}\BibitemShut
  {NoStop}%
\bibitem [{\citenamefont {Kadowaki}\ and\ \citenamefont
  {Nishimori}(1998)}]{KadowakiNishimori1998QuantumAnnealing}%
  \BibitemOpen
  \bibfield  {author} {\bibinfo {author} {\bibfnamefont {T.}~\bibnamefont
  {Kadowaki}}\ and\ \bibinfo {author} {\bibfnamefont {H.}~\bibnamefont
  {Nishimori}},\ }\bibfield  {title} {\emph {\bibinfo {title} {Quantum
  annealing in the transverse Ising model},}\ }\href
  {https://doi.org/10.1103/PhysRevE.58.5355} {\bibfield  {journal} {\bibinfo
  {journal} {Phys. Rev. E}\ }\textbf {\bibinfo {volume} {58}},\ \bibinfo
  {pages} {5355} (\bibinfo {year} {1998})}\BibitemShut {NoStop}%
\bibitem [{\citenamefont {Brooke}\ \emph {et~al.}(1999)\citenamefont {Brooke},
  \citenamefont {Bitko}, \citenamefont {Rosenbaum},\ and\ \citenamefont
  {Aeppli}}]{Brooke99}%
  \BibitemOpen
  \bibfield  {author} {\bibinfo {author} {\bibfnamefont {J.}~\bibnamefont
  {Brooke}}, \bibinfo {author} {\bibfnamefont {D.}~\bibnamefont {Bitko}},
  \bibinfo {author} {\bibfnamefont {T.~F.}\ \bibnamefont {Rosenbaum}},\ and\
  \bibinfo {author} {\bibfnamefont {G.}~\bibnamefont {Aeppli}},\ }\bibfield
  {title} {\emph {\bibinfo {title} {Quantum Annealing of a Disordered
  Magnet},}\ }\href {https://doi.org/10.1126/science.284.5415.779} {\bibfield
  {journal} {\bibinfo  {journal} {Science}\ }\textbf {\bibinfo {volume}
  {284}},\ \bibinfo {pages} {779} (\bibinfo {year} {1999})}\BibitemShut
  {NoStop}%
\bibitem [{\citenamefont {Johnson}\ \emph {et~al.}(2011)\citenamefont
  {Johnson}, \citenamefont {Amin}, \citenamefont {Gildert}, \citenamefont
  {Lanting}, \citenamefont {Hamze}, \citenamefont {Dickson}, \citenamefont
  {Harris}, \citenamefont {Berkley}, \citenamefont {Johansson}, \citenamefont
  {Bunyk}, \citenamefont {Chapple}, \citenamefont {Enderud}, \citenamefont
  {Hilton}, \citenamefont {Karimi}, \citenamefont {Ladizinsky}, \citenamefont
  {Ladizinsky}, \citenamefont {Oh}, \citenamefont {Perminov}, \citenamefont
  {Rich}, \citenamefont {Thom}, \citenamefont {Tolkacheva}, \citenamefont
  {Truncik}, \citenamefont {Uchaikin}, \citenamefont {Wang}, \citenamefont
  {Wilson},\ and\ \citenamefont {Rose}}]{Johnson2011DWave}%
  \BibitemOpen
  \bibfield  {author} {\bibinfo {author} {\bibfnamefont {M.~W.}\ \bibnamefont
  {Johnson}}, \bibinfo {author} {\bibfnamefont {M.~H.~S.}\ \bibnamefont
  {Amin}}, \bibinfo {author} {\bibfnamefont {S.}~\bibnamefont {Gildert}},
  \bibinfo {author} {\bibfnamefont {T.}~\bibnamefont {Lanting}}, \bibinfo
  {author} {\bibfnamefont {F.}~\bibnamefont {Hamze}}, \bibinfo {author}
  {\bibfnamefont {N.}~\bibnamefont {Dickson}}, \bibinfo {author} {\bibfnamefont
  {R.}~\bibnamefont {Harris}}, \bibinfo {author} {\bibfnamefont {A.~J.}\
  \bibnamefont {Berkley}}, \bibinfo {author} {\bibfnamefont {J.}~\bibnamefont
  {Johansson}}, \bibinfo {author} {\bibfnamefont {P.}~\bibnamefont {Bunyk}},
  \bibinfo {author} {\bibfnamefont {E.~M.}\ \bibnamefont {Chapple}}, \bibinfo
  {author} {\bibfnamefont {C.}~\bibnamefont {Enderud}}, \bibinfo {author}
  {\bibfnamefont {J.~P.}\ \bibnamefont {Hilton}}, \bibinfo {author}
  {\bibfnamefont {K.}~\bibnamefont {Karimi}}, \bibinfo {author} {\bibfnamefont
  {E.}~\bibnamefont {Ladizinsky}}, \bibinfo {author} {\bibfnamefont
  {N.}~\bibnamefont {Ladizinsky}}, \bibinfo {author} {\bibfnamefont
  {T.}~\bibnamefont {Oh}}, \bibinfo {author} {\bibfnamefont {I.}~\bibnamefont
  {Perminov}}, \bibinfo {author} {\bibfnamefont {C.}~\bibnamefont {Rich}},
  \bibinfo {author} {\bibfnamefont {M.~C.}\ \bibnamefont {Thom}}, \bibinfo
  {author} {\bibfnamefont {E.}~\bibnamefont {Tolkacheva}}, \bibinfo {author}
  {\bibfnamefont {C.~J.~S.}\ \bibnamefont {Truncik}}, \bibinfo {author}
  {\bibfnamefont {S.}~\bibnamefont {Uchaikin}}, \bibinfo {author}
  {\bibfnamefont {J.}~\bibnamefont {Wang}}, \bibinfo {author} {\bibfnamefont
  {B.}~\bibnamefont {Wilson}},\ and\ \bibinfo {author} {\bibfnamefont
  {G.}~\bibnamefont {Rose}},\ }\bibfield  {title} {\emph {\bibinfo {title}
  {Quantum annealing with manufactured spins},}\ }\href
  {https://doi.org/10.1038/nature10012} {\bibfield  {journal} {\bibinfo
  {journal} {Nature}\ }\textbf {\bibinfo {volume} {473}},\ \bibinfo {pages}
  {194} (\bibinfo {year} {2011})}\BibitemShut {NoStop}%
\bibitem [{\citenamefont {Harris}\ \emph {et~al.}(2010)\citenamefont {Harris},
  \citenamefont {Johnson}, \citenamefont {Lanting}, \citenamefont {Berkley},
  \citenamefont {Johansson}, \citenamefont {Bunyk}, \citenamefont {Tolkacheva},
  \citenamefont {Ladizinsky}, \citenamefont {Ladizinsky}, \citenamefont {Oh},
  \citenamefont {Cioata}, \citenamefont {Perminov}, \citenamefont {Spear},
  \citenamefont {Enderud}, \citenamefont {Rich}, \citenamefont {Uchaikin},
  \citenamefont {Thom}, \citenamefont {Chapple}, \citenamefont {Wang},
  \citenamefont {Wilson}, \citenamefont {Amin}, \citenamefont {Dickson},
  \citenamefont {Karimi}, \citenamefont {Macready}, \citenamefont {Truncik},\
  and\ \citenamefont {Rose}}]{Harris2010DWave}%
  \BibitemOpen
  \bibfield  {author} {\bibinfo {author} {\bibfnamefont {R.}~\bibnamefont
  {Harris}}, \bibinfo {author} {\bibfnamefont {M.~W.}\ \bibnamefont {Johnson}},
  \bibinfo {author} {\bibfnamefont {T.}~\bibnamefont {Lanting}}, \bibinfo
  {author} {\bibfnamefont {A.~J.}\ \bibnamefont {Berkley}}, \bibinfo {author}
  {\bibfnamefont {J.}~\bibnamefont {Johansson}}, \bibinfo {author}
  {\bibfnamefont {P.}~\bibnamefont {Bunyk}}, \bibinfo {author} {\bibfnamefont
  {E.}~\bibnamefont {Tolkacheva}}, \bibinfo {author} {\bibfnamefont
  {E.}~\bibnamefont {Ladizinsky}}, \bibinfo {author} {\bibfnamefont
  {N.}~\bibnamefont {Ladizinsky}}, \bibinfo {author} {\bibfnamefont
  {T.}~\bibnamefont {Oh}}, \bibinfo {author} {\bibfnamefont {F.}~\bibnamefont
  {Cioata}}, \bibinfo {author} {\bibfnamefont {I.}~\bibnamefont {Perminov}},
  \bibinfo {author} {\bibfnamefont {P.}~\bibnamefont {Spear}}, \bibinfo
  {author} {\bibfnamefont {C.}~\bibnamefont {Enderud}}, \bibinfo {author}
  {\bibfnamefont {C.}~\bibnamefont {Rich}}, \bibinfo {author} {\bibfnamefont
  {S.}~\bibnamefont {Uchaikin}}, \bibinfo {author} {\bibfnamefont {M.~C.}\
  \bibnamefont {Thom}}, \bibinfo {author} {\bibfnamefont {E.~M.}\ \bibnamefont
  {Chapple}}, \bibinfo {author} {\bibfnamefont {J.}~\bibnamefont {Wang}},
  \bibinfo {author} {\bibfnamefont {B.}~\bibnamefont {Wilson}}, \bibinfo
  {author} {\bibfnamefont {M.~H.~S.}\ \bibnamefont {Amin}}, \bibinfo {author}
  {\bibfnamefont {N.}~\bibnamefont {Dickson}}, \bibinfo {author} {\bibfnamefont
  {K.}~\bibnamefont {Karimi}}, \bibinfo {author} {\bibfnamefont
  {B.}~\bibnamefont {Macready}}, \bibinfo {author} {\bibfnamefont {C.~J.~S.}\
  \bibnamefont {Truncik}},\ and\ \bibinfo {author} {\bibfnamefont
  {G.}~\bibnamefont {Rose}},\ }\bibfield  {title} {\emph {\bibinfo {title}
  {Experimental investigation of an eight-qubit unit cell in a superconducting
  optimization processor},}\ }\href
  {https://doi.org/10.1103/PhysRevB.82.024511} {\bibfield  {journal} {\bibinfo
  {journal} {Phys. Rev. B}\ }\textbf {\bibinfo {volume} {82}},\ \bibinfo
  {pages} {024511} (\bibinfo {year} {2010})}\BibitemShut {NoStop}%
\bibitem [{\citenamefont {Bunyk}\ \emph {et~al.}(2014)\citenamefont {Bunyk},
  \citenamefont {Hoskinson}, \citenamefont {Johnson}, \citenamefont
  {Tolkacheva}, \citenamefont {Altomare}, \citenamefont {Berkley},
  \citenamefont {Harris}, \citenamefont {Hilton}, \citenamefont {Lanting},
  \citenamefont {Przybysz},\ and\ \citenamefont {Whittaker}}]{Bunyk2014DWave}%
  \BibitemOpen
  \bibfield  {author} {\bibinfo {author} {\bibfnamefont {P.~I.}\ \bibnamefont
  {Bunyk}}, \bibinfo {author} {\bibfnamefont {E.~M.}\ \bibnamefont
  {Hoskinson}}, \bibinfo {author} {\bibfnamefont {M.~W.}\ \bibnamefont
  {Johnson}}, \bibinfo {author} {\bibfnamefont {E.}~\bibnamefont {Tolkacheva}},
  \bibinfo {author} {\bibfnamefont {F.}~\bibnamefont {Altomare}}, \bibinfo
  {author} {\bibfnamefont {A.~J.}\ \bibnamefont {Berkley}}, \bibinfo {author}
  {\bibfnamefont {R.}~\bibnamefont {Harris}}, \bibinfo {author} {\bibfnamefont
  {J.~P.}\ \bibnamefont {Hilton}}, \bibinfo {author} {\bibfnamefont
  {T.}~\bibnamefont {Lanting}}, \bibinfo {author} {\bibfnamefont {A.~J.}\
  \bibnamefont {Przybysz}},\ and\ \bibinfo {author} {\bibfnamefont
  {J.}~\bibnamefont {Whittaker}},\ }\bibfield  {title} {\emph {\bibinfo {title}
  {Architectural Considerations in the Design of a Superconducting Quantum
  Annealing Processor},}\ }\href {https://doi.org/10.1109/TASC.2014.2318294}
  {\bibfield  {journal} {\bibinfo  {journal} {{IEEE} Transactions on Applied
  Superconductivity}\ }\textbf {\bibinfo {volume} {24}},\ \bibinfo {pages} {1}
  (\bibinfo {year} {2014})}\BibitemShut {NoStop}%
\bibitem [{\citenamefont {Albash}\ and\ \citenamefont
  {Lidar}(2018{\natexlab{a}})}]{AlbashLidar2018AdiabaticQuantumComputation}%
  \BibitemOpen
  \bibfield  {author} {\bibinfo {author} {\bibfnamefont {T.}~\bibnamefont
  {Albash}}\ and\ \bibinfo {author} {\bibfnamefont {D.~A.}\ \bibnamefont
  {Lidar}},\ }\bibfield  {title} {\emph {\bibinfo {title} {Adiabatic quantum
  computation},}\ }\href {https://doi.org/10.1103/RevModPhys.90.015002}
  {\bibfield  {journal} {\bibinfo  {journal} {Rev. Mod. Phys.}\ }\textbf
  {\bibinfo {volume} {90}},\ \bibinfo {pages} {015002} (\bibinfo {year}
  {2018}{\natexlab{a}})}\BibitemShut {NoStop}%
\bibitem [{\citenamefont {Albash}\ and\ \citenamefont
  {Lidar}(2018{\natexlab{b}})}]{Albash2018QuantumAnnealingScalingAdvantageOverSimulatedAnnealing}%
  \BibitemOpen
  \bibfield  {author} {\bibinfo {author} {\bibfnamefont {T.}~\bibnamefont
  {Albash}}\ and\ \bibinfo {author} {\bibfnamefont {D.~A.}\ \bibnamefont
  {Lidar}},\ }\bibfield  {title} {\emph {\bibinfo {title} {Demonstration of a
  Scaling Advantage for a Quantum Annealer over Simulated Annealing},}\ }\href
  {https://doi.org/10.1103/PhysRevX.8.031016} {\bibfield  {journal} {\bibinfo
  {journal} {Phys. Rev. X}\ }\textbf {\bibinfo {volume} {8}},\ \bibinfo {pages}
  {031016} (\bibinfo {year} {2018}{\natexlab{b}})}\BibitemShut {NoStop}%
\bibitem [{\citenamefont {King}\ \emph {et~al.}(2023)\citenamefont {King},
  \citenamefont {Raymond}, \citenamefont {Lanting}, \citenamefont {Harris},
  \citenamefont {Zucca}, \citenamefont {Altomare}, \citenamefont {Berkley},
  \citenamefont {Boothby}, \citenamefont {Ejtemaee}, \citenamefont {Enderud},
  \citenamefont {Hoskinson}, \citenamefont {Huang}, \citenamefont {Ladizinsky},
  \citenamefont {MacDonald}, \citenamefont {Marsden}, \citenamefont {Molavi},
  \citenamefont {Oh}, \citenamefont {Poulin-Lamarre}, \citenamefont {Reis},
  \citenamefont {Rich}, \citenamefont {Sato}, \citenamefont {Tsai},
  \citenamefont {Volkmann}, \citenamefont {Whittaker}, \citenamefont {Yao},
  \citenamefont {Sandvik},\ and\ \citenamefont
  {Amin}}]{King2023QuantumCriticalDynamics5000SpinGlass}%
  \BibitemOpen
  \bibfield  {author} {\bibinfo {author} {\bibfnamefont {A.~D.}\ \bibnamefont
  {King}}, \bibinfo {author} {\bibfnamefont {J.}~\bibnamefont {Raymond}},
  \bibinfo {author} {\bibfnamefont {T.}~\bibnamefont {Lanting}}, \bibinfo
  {author} {\bibfnamefont {R.}~\bibnamefont {Harris}}, \bibinfo {author}
  {\bibfnamefont {A.}~\bibnamefont {Zucca}}, \bibinfo {author} {\bibfnamefont
  {F.}~\bibnamefont {Altomare}}, \bibinfo {author} {\bibfnamefont {A.~J.}\
  \bibnamefont {Berkley}}, \bibinfo {author} {\bibfnamefont {K.}~\bibnamefont
  {Boothby}}, \bibinfo {author} {\bibfnamefont {S.}~\bibnamefont {Ejtemaee}},
  \bibinfo {author} {\bibfnamefont {C.}~\bibnamefont {Enderud}}, \bibinfo
  {author} {\bibfnamefont {E.}~\bibnamefont {Hoskinson}}, \bibinfo {author}
  {\bibfnamefont {S.}~\bibnamefont {Huang}}, \bibinfo {author} {\bibfnamefont
  {E.}~\bibnamefont {Ladizinsky}}, \bibinfo {author} {\bibfnamefont {A.~J.~R.}\
  \bibnamefont {MacDonald}}, \bibinfo {author} {\bibfnamefont {G.}~\bibnamefont
  {Marsden}}, \bibinfo {author} {\bibfnamefont {R.}~\bibnamefont {Molavi}},
  \bibinfo {author} {\bibfnamefont {T.}~\bibnamefont {Oh}}, \bibinfo {author}
  {\bibfnamefont {G.}~\bibnamefont {Poulin-Lamarre}}, \bibinfo {author}
  {\bibfnamefont {M.}~\bibnamefont {Reis}}, \bibinfo {author} {\bibfnamefont
  {C.}~\bibnamefont {Rich}}, \bibinfo {author} {\bibfnamefont {Y.}~\bibnamefont
  {Sato}}, \bibinfo {author} {\bibfnamefont {N.}~\bibnamefont {Tsai}}, \bibinfo
  {author} {\bibfnamefont {M.}~\bibnamefont {Volkmann}}, \bibinfo {author}
  {\bibfnamefont {J.~D.}\ \bibnamefont {Whittaker}}, \bibinfo {author}
  {\bibfnamefont {J.}~\bibnamefont {Yao}}, \bibinfo {author} {\bibfnamefont
  {A.~W.}\ \bibnamefont {Sandvik}},\ and\ \bibinfo {author} {\bibfnamefont
  {M.~H.}\ \bibnamefont {Amin}},\ }\bibfield  {title} {\emph {\bibinfo {title}
  {Quantum critical dynamics in a 5,000-qubit programmable spin glass},}\
  }\href {https://doi.org/10.1038/s41586-023-05867-2} {\bibfield  {journal}
  {\bibinfo  {journal} {Nature}\ }\textbf {\bibinfo {volume} {617}},\ \bibinfo
  {pages} {61} (\bibinfo {year} {2023})}\BibitemShut {NoStop}%
\bibitem [{\citenamefont {Callison}\ \emph {et~al.}(2019)\citenamefont
  {Callison}, \citenamefont {Chancellor}, \citenamefont {Mintert},\ and\
  \citenamefont {Kendon}}]{FindingSpinglas_Callison}%
  \BibitemOpen
  \bibfield  {author} {\bibinfo {author} {\bibfnamefont {A.}~\bibnamefont
  {Callison}}, \bibinfo {author} {\bibfnamefont {N.}~\bibnamefont
  {Chancellor}}, \bibinfo {author} {\bibfnamefont {F.}~\bibnamefont
  {Mintert}},\ and\ \bibinfo {author} {\bibfnamefont {V.}~\bibnamefont
  {Kendon}},\ }\bibfield  {title} {\emph {\bibinfo {title} {Finding spin glass
  ground states using quantum walks},}\ }\href
  {https://doi.org/10.1088/1367-2630/ab5ca2} {\bibfield  {journal} {\bibinfo
  {journal} {New Journal of Physics}\ }\textbf {\bibinfo {volume} {21}},\
  \bibinfo {pages} {123022} (\bibinfo {year} {2019})}\BibitemShut {NoStop}%
\bibitem [{\citenamefont
  {Neuhaus}(2014{\natexlab{a}})}]{neuhaus2014monteCarloQA2SAT}%
  \BibitemOpen
  \bibfield  {author} {\bibinfo {author} {\bibfnamefont {T.}~\bibnamefont
  {Neuhaus}},\ }\bibfield  {title} {\emph {\bibinfo {title} {Monte Carlo Search
  for Very Hard KSAT Realizations for Use in Quantum Annealing},}\ }\Eprint
  {https://arxiv.org/abs/1412.5361} {arXiv:1412.5361 [cond-mat.stat-mech]}
  (\bibinfo {year} {2014}{\natexlab{a}})\BibitemShut {NoStop}%
\bibitem [{\citenamefont
  {Neuhaus}(2014{\natexlab{b}})}]{neuhaus2014quantumSearches2SAT}%
  \BibitemOpen
  \bibfield  {author} {\bibinfo {author} {\bibfnamefont {T.}~\bibnamefont
  {Neuhaus}},\ }\bibfield  {title} {\emph {\bibinfo {title} {Quantum Searches
  in a Hard 2SAT Ensemble},}\ }\Eprint {https://arxiv.org/abs/1412.5460}
  {arXiv:1412.5460 [quant-ph]}  (\bibinfo {year}
  {2014}{\natexlab{b}})\BibitemShut {NoStop}%
\bibitem [{\citenamefont {Mehta}\ \emph {et~al.}(2021)\citenamefont {Mehta},
  \citenamefont {Jin}, \citenamefont {{De Raedt}},\ and\ \citenamefont
  {Michielsen}}]{Mehta2021quantumAnnealingWithTriggerHamiltonians2SATNonstoquastic}%
  \BibitemOpen
  \bibfield  {author} {\bibinfo {author} {\bibfnamefont {V.}~\bibnamefont
  {Mehta}}, \bibinfo {author} {\bibfnamefont {F.}~\bibnamefont {Jin}}, \bibinfo
  {author} {\bibfnamefont {H.}~\bibnamefont {{De Raedt}}},\ and\ \bibinfo
  {author} {\bibfnamefont {K.}~\bibnamefont {Michielsen}},\ }\bibfield  {title}
  {\emph {\bibinfo {title} {Quantum annealing with trigger Hamiltonians:
  Application to 2-satisfiability and nonstoquastic problems},}\ }\href
  {https://doi.org/10.1103/PhysRevA.104.032421} {\bibfield  {journal} {\bibinfo
   {journal} {Phys. Rev. A}\ }\textbf {\bibinfo {volume} {104}},\ \bibinfo
  {pages} {032421} (\bibinfo {year} {2021})}\BibitemShut {NoStop}%
\bibitem [{\citenamefont {Mehta}\ \emph
  {et~al.}(2022{\natexlab{a}})\citenamefont {Mehta}, \citenamefont {Jin},
  \citenamefont {{De Raedt}},\ and\ \citenamefont
  {Michielsen}}]{mehta2022QuantumAnnealingForHard2SATProblems}%
  \BibitemOpen
  \bibfield  {author} {\bibinfo {author} {\bibfnamefont {V.}~\bibnamefont
  {Mehta}}, \bibinfo {author} {\bibfnamefont {F.}~\bibnamefont {Jin}}, \bibinfo
  {author} {\bibfnamefont {H.}~\bibnamefont {{De Raedt}}},\ and\ \bibinfo
  {author} {\bibfnamefont {K.}~\bibnamefont {Michielsen}},\ }\bibfield  {title}
  {\emph {\bibinfo {title} {Quantum annealing for hard 2-satisfiability
  problems: Distribution and scaling of minimum energy gap and success
  probability},}\ }\href {https://doi.org/10.1103/PhysRevA.105.062406}
  {\bibfield  {journal} {\bibinfo  {journal} {Phys. Rev. A}\ }\textbf {\bibinfo
  {volume} {105}},\ \bibinfo {pages} {062406} (\bibinfo {year}
  {2022}{\natexlab{a}})}\BibitemShut {NoStop}%
\bibitem [{\citenamefont {Mehta}\ \emph
  {et~al.}(2022{\natexlab{b}})\citenamefont {Mehta}, \citenamefont {Jin},
  \citenamefont {Michielsen},\ and\ \citenamefont {{De
  Raedt}}}]{mehta2022hardnessOfQUBO}%
  \BibitemOpen
  \bibfield  {author} {\bibinfo {author} {\bibfnamefont {V.}~\bibnamefont
  {Mehta}}, \bibinfo {author} {\bibfnamefont {F.}~\bibnamefont {Jin}}, \bibinfo
  {author} {\bibfnamefont {K.}~\bibnamefont {Michielsen}},\ and\ \bibinfo
  {author} {\bibfnamefont {H.}~\bibnamefont {{De Raedt}}},\ }\bibfield  {title}
  {\emph {\bibinfo {title} {On the hardness of quadratic unconstrained binary
  optimization problems},}\ }\href {https://doi.org/10.3389/fphy.2022.956882}
  {\bibfield  {journal} {\bibinfo  {journal} {Front. Phys.}\ }\textbf {\bibinfo
  {volume} {10}},\ \bibinfo {pages} {956882} (\bibinfo {year}
  {2022}{\natexlab{b}})}\BibitemShut {NoStop}%
\bibitem [{\citenamefont {Morley}\ \emph {et~al.}(2019)\citenamefont {Morley},
  \citenamefont {Chancellor}, \citenamefont {Bose},\ and\ \citenamefont
  {Kendon}}]{HybridQAQW_Morley}%
  \BibitemOpen
  \bibfield  {author} {\bibinfo {author} {\bibfnamefont {J.~G.}\ \bibnamefont
  {Morley}}, \bibinfo {author} {\bibfnamefont {N.}~\bibnamefont {Chancellor}},
  \bibinfo {author} {\bibfnamefont {S.}~\bibnamefont {Bose}},\ and\ \bibinfo
  {author} {\bibfnamefont {V.}~\bibnamefont {Kendon}},\ }\bibfield  {title}
  {\emph {\bibinfo {title} {Quantum search with hybrid adiabatic--quantum-walk
  algorithms and realistic noise},}\ }\href
  {https://doi.org/10.1103/PhysRevA.99.022339} {\bibfield  {journal} {\bibinfo
  {journal} {Phys. Rev. A}\ }\textbf {\bibinfo {volume} {99}},\ \bibinfo
  {pages} {022339} (\bibinfo {year} {2019})}\BibitemShut {NoStop}%
\bibitem [{\citenamefont {{Lishan Zeng, Jun Zhang and Mohan
  Sarovar}}(2016)}]{QAOpt_PathOptZeng}%
  \BibitemOpen
  \bibfield  {author} {\bibinfo {author} {\bibnamefont {{Lishan Zeng, Jun Zhang
  and Mohan Sarovar}}},\ }\bibfield  {title} {\emph {\bibinfo {title}
  {{Schedule path optimization for adiabatic quantum computing and
  optimization}},}\ }\bibfield  {journal} {\bibinfo  {journal} {Journal of
  Physics A: Mathematical and Theoretical}\ }\href
  {https://doi.org/10.1088/1751-8113/49/16/165305}
  {10.1088/1751-8113/49/16/165305} (\bibinfo {year} {2016})\BibitemShut
  {NoStop}%
\bibitem [{\citenamefont {Herr}\ \emph {et~al.}(2017)\citenamefont {Herr},
  \citenamefont {Brown}, \citenamefont {Heim}, \citenamefont {Könz},
  \citenamefont {Mazzola},\ and\ \citenamefont {Troyer}}]{QAOpt_PathOptHerr}%
  \BibitemOpen
  \bibfield  {author} {\bibinfo {author} {\bibfnamefont {D.}~\bibnamefont
  {Herr}}, \bibinfo {author} {\bibfnamefont {E.}~\bibnamefont {Brown}},
  \bibinfo {author} {\bibfnamefont {B.}~\bibnamefont {Heim}}, \bibinfo {author}
  {\bibfnamefont {M.}~\bibnamefont {Könz}}, \bibinfo {author} {\bibfnamefont
  {G.}~\bibnamefont {Mazzola}},\ and\ \bibinfo {author} {\bibfnamefont
  {M.}~\bibnamefont {Troyer}},\ }\bibfield  {title} {\emph {\bibinfo {title}
  {Optimizing Schedules for Quantum Annealing},}\ }\Eprint
  {https://arxiv.org/abs/1705.00420} {arXiv:1705.00420}  (\bibinfo {year}
  {2017})\BibitemShut {NoStop}%
\bibitem [{\citenamefont {{Yu-Qin Chen and Yu Chen and Chee-Kong Lee and
  Shengyu Zhang and Chang-Yu Hsieh}}(2022)}]{QAOpt_MonteCarloTreeSearch}%
  \BibitemOpen
  \bibfield  {author} {\bibinfo {author} {\bibnamefont {{Yu-Qin Chen and Yu
  Chen and Chee-Kong Lee and Shengyu Zhang and Chang-Yu Hsieh}}},\ }\bibfield
  {title} {\emph {\bibinfo {title} {{Optimizing Quantum Annealing Schedules
  with Monte Carlo Tree Search enhanced with neural networks}},}\ }\bibfield
  {journal} {\bibinfo  {journal} {Nature Machine Intelligence}\ }\href
  {https://doi.org/10.1038/s42256-022-00446-y} {10.1038/s42256-022-00446-y}
  (\bibinfo {year} {2022})\BibitemShut {NoStop}%
\bibitem [{\citenamefont {Karp}(1972)}]{Karp1972KarpsNPCompleteProblems}%
  \BibitemOpen
  \bibfield  {author} {\bibinfo {author} {\bibfnamefont {R.~M.}\ \bibnamefont
  {Karp}},\ }\emph {\bibinfo {title} {Reducibility among Combinatorial
  Problems},}\ in\ \href {https://doi.org/10.1007/978-1-4684-2001-2_9} {\emph
  {\bibinfo {booktitle} {Complexity of Computer Computations: Proceedings of a
  symposium on the Complexity of Computer Computations, held March 20--22,
  1972, at the IBM Thomas J. Watson Research Center, Yorktown Heights, New
  York, and sponsored by the Office of Naval Research, Mathematics Program, IBM
  World Trade Corporation, and the IBM Research Mathematical Sciences
  Department}}},\ \bibinfo {editor} {edited by\ \bibinfo {editor}
  {\bibfnamefont {R.~E.}\ \bibnamefont {Miller}}, \bibinfo {editor}
  {\bibfnamefont {J.~W.}\ \bibnamefont {Thatcher}},\ and\ \bibinfo {editor}
  {\bibfnamefont {J.~D.}\ \bibnamefont {Bohlinger}}}\ (\bibinfo  {publisher}
  {Springer US},\ \bibinfo {address} {Boston, MA},\ \bibinfo {year} {1972})\
  pp.\ \bibinfo {pages} {85--103}\BibitemShut {NoStop}%
\bibitem [{\citenamefont {Choi}(2010)}]{EC_Choi}%
  \BibitemOpen
  \bibfield  {author} {\bibinfo {author} {\bibfnamefont {V.}~\bibnamefont
  {Choi}},\ }\href@noop {} {\emph {\bibinfo {title} {Adiabatic Quantum
  Algorithms for the NP-Complete Maximum-Weight Independent Set, Exact Cover
  and 3SAT Problems},}\ } (\bibinfo {year} {2010}),\ \Eprint
  {https://arxiv.org/abs/1004.2226} {arXiv:1004.2226 [quant-ph]} \BibitemShut
  {NoStop}%
\bibitem [{\citenamefont {Dantzig}\ \emph {et~al.}(1954)\citenamefont
  {Dantzig}, \citenamefont {Fulkerson},\ and\ \citenamefont
  {Johnson}}]{Dantzig1954SolutionOfALargeScaleTSP}%
  \BibitemOpen
  \bibfield  {author} {\bibinfo {author} {\bibfnamefont {G.}~\bibnamefont
  {Dantzig}}, \bibinfo {author} {\bibfnamefont {R.}~\bibnamefont {Fulkerson}},\
  and\ \bibinfo {author} {\bibfnamefont {S.}~\bibnamefont {Johnson}},\
  }\bibfield  {title} {\emph {\bibinfo {title} {Solution of a Large-Scale
  Traveling-Salesman Problem},}\ }\href {https://doi.org/10.1287/opre.2.4.393}
  {\bibfield  {journal} {\bibinfo  {journal} {J. Oper. Res. Soc. Am.}\ }\textbf
  {\bibinfo {volume} {2}},\ \bibinfo {pages} {393} (\bibinfo {year}
  {1954})}\BibitemShut {NoStop}%
\bibitem [{\citenamefont {Bellman}(1962)}]{Bellman1962TSPDynamicProgramming}%
  \BibitemOpen
  \bibfield  {author} {\bibinfo {author} {\bibfnamefont {R.}~\bibnamefont
  {Bellman}},\ }\bibfield  {title} {\emph {\bibinfo {title} {Dynamic
  Programming Treatment of the Travelling Salesman Problem},}\ }\href
  {https://doi.org/10.1145/321105.321111} {\bibfield  {journal} {\bibinfo
  {journal} {J. ACM}\ }\textbf {\bibinfo {volume} {9}},\ \bibinfo {pages} {61}
  (\bibinfo {year} {1962})}\BibitemShut {NoStop}%
\bibitem [{\citenamefont {Held}\ and\ \citenamefont
  {Karp}(1962)}]{HeldKarp1962TSP}%
  \BibitemOpen
  \bibfield  {author} {\bibinfo {author} {\bibfnamefont {M.}~\bibnamefont
  {Held}}\ and\ \bibinfo {author} {\bibfnamefont {R.~M.}\ \bibnamefont
  {Karp}},\ }\bibfield  {title} {\emph {\bibinfo {title} {A Dynamic Programming
  Approach to Sequencing Problems},}\ }\href {https://doi.org/10.1137/0110015}
  {\bibfield  {journal} {\bibinfo  {journal} {Journal of the Society for
  Industrial and Applied Mathematics}\ }\textbf {\bibinfo {volume} {10}},\
  \bibinfo {pages} {196} (\bibinfo {year} {1962})}\BibitemShut {NoStop}%
\bibitem [{\citenamefont
  {Lucas}(2014)}]{Lucas2014IsingQUBOFormulationManyNPproblems}%
  \BibitemOpen
  \bibfield  {author} {\bibinfo {author} {\bibfnamefont {A.}~\bibnamefont
  {Lucas}},\ }\bibfield  {title} {\emph {\bibinfo {title} {Ising formulations
  of many NP problems},}\ }\href {https://doi.org/10.3389/fphy.2014.00005}
  {\bibfield  {journal} {\bibinfo  {journal} {Front. Phys.}\ }\textbf {\bibinfo
  {volume} {2}},\ \bibinfo {pages} {5} (\bibinfo {year} {2014})}\BibitemShut
  {NoStop}%
\bibitem [{\citenamefont {Gonzalez~Calaza}\ \emph {et~al.}(2021)\citenamefont
  {Gonzalez~Calaza}, \citenamefont {Willsch},\ and\ \citenamefont
  {Michielsen}}]{calaza2021gardenoptimization}%
  \BibitemOpen
  \bibfield  {author} {\bibinfo {author} {\bibfnamefont {C.~D.}\ \bibnamefont
  {Gonzalez~Calaza}}, \bibinfo {author} {\bibfnamefont {D.}~\bibnamefont
  {Willsch}},\ and\ \bibinfo {author} {\bibfnamefont {K.}~\bibnamefont
  {Michielsen}},\ }\bibfield  {title} {\emph {\bibinfo {title} {Garden
  optimization problems for benchmarking quantum annealers},}\ }\href
  {https://doi.org/10.1007/s11128-021-03226-6} {\bibfield  {journal} {\bibinfo
  {journal} {Quantum Inf. Process.}\ }\textbf {\bibinfo {volume} {20}},\
  \bibinfo {pages} {305} (\bibinfo {year} {2021})}\BibitemShut {NoStop}%
\bibitem [{\citenamefont {{J\"{u}lich Supercomputing
  Centre}}(2021)}]{JuwelsClusterBooster}%
  \BibitemOpen
  \bibfield  {author} {\bibinfo {author} {\bibnamefont {{J\"{u}lich
  Supercomputing Centre}}},\ }\bibfield  {title} {\emph {\bibinfo {title}
  {{JUWELS Cluster and Booster: Exascale Pathfinder with Modular Supercomputing
  Architecture at Juelich Supercomputing Centre}},}\ }\href
  {https://doi.org/10.17815/jlsrf-7-183} {\bibfield  {journal} {\bibinfo
  {journal} {J. of Large-Scale Res. Facil.}\ }\textbf {\bibinfo {volume} {7}},\
  \bibinfo {pages} {A138} (\bibinfo {year} {2021})}\BibitemShut {NoStop}%
\bibitem [{\citenamefont {Grover}(1996)}]{Grover1996Algorithm}%
  \BibitemOpen
  \bibfield  {author} {\bibinfo {author} {\bibfnamefont {L.~K.}\ \bibnamefont
  {Grover}},\ }\bibfield  {title} {\emph {\bibinfo {title} {A Fast Quantum
  Mechanical Algorithm for Database Search},}\ }in\ \href
  {https://doi.org/10.1145/237814.237866} {\emph {\bibinfo {booktitle}
  {Proceedings of the Twenty-Eighth Annual ACM Symposium on Theory of
  Computing}}},\ \bibinfo {series and number} {STOC '96}\ (\bibinfo
  {publisher} {Association for Computing Machinery},\ \bibinfo {address} {New
  York, NY, USA},\ \bibinfo {year} {1996})\ p.\ \bibinfo {pages}
  {212–219}\BibitemShut {NoStop}%
\bibitem [{\citenamefont {Farhi}\ \emph {et~al.}(2014)\citenamefont {Farhi},
  \citenamefont {Goldstone},\ and\ \citenamefont {Gutmann}}]{Farhi2014QAOA}%
  \BibitemOpen
  \bibfield  {author} {\bibinfo {author} {\bibfnamefont {E.}~\bibnamefont
  {Farhi}}, \bibinfo {author} {\bibfnamefont {J.}~\bibnamefont {Goldstone}},\
  and\ \bibinfo {author} {\bibfnamefont {S.}~\bibnamefont {Gutmann}},\
  }\bibfield  {title} {\emph {\bibinfo {title} {A Quantum Approximate
  Optimization Algorithm},}\ }\Eprint {https://arxiv.org/abs/1411.4028}
  {arXiv:1411.4028 [quant-ph]}  (\bibinfo {year} {2014})\BibitemShut {NoStop}%
\bibitem [{\citenamefont {McClean}\ \emph {et~al.}(2016)\citenamefont
  {McClean}, \citenamefont {Romero}, \citenamefont {Babbush},\ and\
  \citenamefont
  {Aspuru-Guzik}}]{McClean2016TheoryVariationalQuantumEigensolver}%
  \BibitemOpen
  \bibfield  {author} {\bibinfo {author} {\bibfnamefont {J.~R.}\ \bibnamefont
  {McClean}}, \bibinfo {author} {\bibfnamefont {J.}~\bibnamefont {Romero}},
  \bibinfo {author} {\bibfnamefont {R.}~\bibnamefont {Babbush}},\ and\ \bibinfo
  {author} {\bibfnamefont {A.}~\bibnamefont {Aspuru-Guzik}},\ }\bibfield
  {title} {\emph {\bibinfo {title} {The theory of variational hybrid
  quantum-classical algorithms},}\ }\href
  {https://doi.org/10.1088/1367-2630/18/2/023023} {\bibfield  {journal}
  {\bibinfo  {journal} {New J. Phys.}\ }\textbf {\bibinfo {volume} {18}},\
  \bibinfo {pages} {023023} (\bibinfo {year} {2016})}\BibitemShut {NoStop}%
\bibitem [{\citenamefont {Tilly}\ \emph {et~al.}(2022)\citenamefont {Tilly},
  \citenamefont {Chen}, \citenamefont {Cao}, \citenamefont {Picozzi},
  \citenamefont {Setia}, \citenamefont {Li}, \citenamefont {Grant},
  \citenamefont {Wossnig}, \citenamefont {Rungger}, \citenamefont {Booth},\
  and\ \citenamefont {Tennyson}}]{Tilly20221VQE}%
  \BibitemOpen
  \bibfield  {author} {\bibinfo {author} {\bibfnamefont {J.}~\bibnamefont
  {Tilly}}, \bibinfo {author} {\bibfnamefont {H.}~\bibnamefont {Chen}},
  \bibinfo {author} {\bibfnamefont {S.}~\bibnamefont {Cao}}, \bibinfo {author}
  {\bibfnamefont {D.}~\bibnamefont {Picozzi}}, \bibinfo {author} {\bibfnamefont
  {K.}~\bibnamefont {Setia}}, \bibinfo {author} {\bibfnamefont
  {Y.}~\bibnamefont {Li}}, \bibinfo {author} {\bibfnamefont {E.}~\bibnamefont
  {Grant}}, \bibinfo {author} {\bibfnamefont {L.}~\bibnamefont {Wossnig}},
  \bibinfo {author} {\bibfnamefont {I.}~\bibnamefont {Rungger}}, \bibinfo
  {author} {\bibfnamefont {G.~H.}\ \bibnamefont {Booth}},\ and\ \bibinfo
  {author} {\bibfnamefont {J.}~\bibnamefont {Tennyson}},\ }\bibfield  {title}
  {\emph {\bibinfo {title} {The Variational Quantum Eigensolver: A review of
  methods and best practices},}\ }\href
  {https://doi.org/https://doi.org/10.1016/j.physrep.2022.08.003} {\bibfield
  {journal} {\bibinfo  {journal} {Phys. Rep.}\ }\textbf {\bibinfo {volume}
  {986}},\ \bibinfo {pages} {1} (\bibinfo {year} {2022})}\BibitemShut {NoStop}%
\bibitem [{\citenamefont {Farhi}\ \emph {et~al.}(2000)\citenamefont {Farhi},
  \citenamefont {Goldstone}, \citenamefont {Gutmann},\ and\ \citenamefont
  {Sipser}}]{Farhi2000AdiabaticQuantumComputation}%
  \BibitemOpen
  \bibfield  {author} {\bibinfo {author} {\bibfnamefont {E.}~\bibnamefont
  {Farhi}}, \bibinfo {author} {\bibfnamefont {J.}~\bibnamefont {Goldstone}},
  \bibinfo {author} {\bibfnamefont {S.}~\bibnamefont {Gutmann}},\ and\ \bibinfo
  {author} {\bibfnamefont {M.}~\bibnamefont {Sipser}},\ }\bibfield  {title}
  {\emph {\bibinfo {title} {Quantum Computation by Adiabatic Evolution},}\
  }\Eprint {https://arxiv.org/abs/quant-ph/0001106} {arXiv:quant-ph/0001106}
  (\bibinfo {year} {2000})\BibitemShut {NoStop}%
\bibitem [{\citenamefont {Zhou}\ \emph {et~al.}(2020)\citenamefont {Zhou},
  \citenamefont {Wang}, \citenamefont {Choi}, \citenamefont {Pichler},\ and\
  \citenamefont {Lukin}}]{QAOA_Opt_Zhou}%
  \BibitemOpen
  \bibfield  {author} {\bibinfo {author} {\bibfnamefont {L.}~\bibnamefont
  {Zhou}}, \bibinfo {author} {\bibfnamefont {S.-T.}\ \bibnamefont {Wang}},
  \bibinfo {author} {\bibfnamefont {S.}~\bibnamefont {Choi}}, \bibinfo {author}
  {\bibfnamefont {H.}~\bibnamefont {Pichler}},\ and\ \bibinfo {author}
  {\bibfnamefont {M.~D.}\ \bibnamefont {Lukin}},\ }\bibfield  {title} {\emph
  {\bibinfo {title} {Quantum Approximate Optimization Algorithm: Performance,
  Mechanism, and Implementation on Near-Term Devices},}\ }\href
  {https://doi.org/10.1103/PhysRevX.10.021067} {\bibfield  {journal} {\bibinfo
  {journal} {Phys. Rev. X}\ }\textbf {\bibinfo {volume} {10}},\ \bibinfo
  {pages} {021067} (\bibinfo {year} {2020})}\BibitemShut {NoStop}%
\bibitem [{\citenamefont {Sack}\ and\ \citenamefont
  {Serbyn}(2021)}]{QAOA_Opt_Sack}%
  \BibitemOpen
  \bibfield  {author} {\bibinfo {author} {\bibfnamefont {S.~H.}\ \bibnamefont
  {Sack}}\ and\ \bibinfo {author} {\bibfnamefont {M.}~\bibnamefont {Serbyn}},\
  }\bibfield  {title} {\emph {\bibinfo {title} {Quantum annealing
  initialization of the quantum approximate optimization algorithm},}\ }\href
  {https://doi.org/10.22331/q-2021-07-01-491} {\bibfield  {journal} {\bibinfo
  {journal} {Quantum}\ }\textbf {\bibinfo {volume} {5}},\ \bibinfo {pages}
  {491} (\bibinfo {year} {2021})}\BibitemShut {NoStop}%
\bibitem [{\citenamefont {Nelder}\ and\ \citenamefont
  {Mead}(1965)}]{NelderMead1965}%
  \BibitemOpen
  \bibfield  {author} {\bibinfo {author} {\bibfnamefont {J.~A.}\ \bibnamefont
  {Nelder}}\ and\ \bibinfo {author} {\bibfnamefont {R.}~\bibnamefont {Mead}},\
  }\bibfield  {title} {\emph {\bibinfo {title} {A Simplex Method for Function
  Minimization},}\ }\href {https://doi.org/10.1093/comjnl/7.4.308} {\bibfield
  {journal} {\bibinfo  {journal} {Comput. J.}\ }\textbf {\bibinfo {volume}
  {7}},\ \bibinfo {pages} {308} (\bibinfo {year} {1965})}\BibitemShut {NoStop}%
\bibitem [{\citenamefont {Press}\ \emph {et~al.}(2007)\citenamefont {Press},
  \citenamefont {Teukolsky}, \citenamefont {Vetterling},\ and\ \citenamefont
  {Flannery}}]{numericalrecipes}%
  \BibitemOpen
  \bibfield  {author} {\bibinfo {author} {\bibfnamefont {W.~H.}\ \bibnamefont
  {Press}}, \bibinfo {author} {\bibfnamefont {S.~A.}\ \bibnamefont
  {Teukolsky}}, \bibinfo {author} {\bibfnamefont {W.~T.}\ \bibnamefont
  {Vetterling}},\ and\ \bibinfo {author} {\bibfnamefont {B.~P.}\ \bibnamefont
  {Flannery}},\ }\href {http://numerical.recipes/} {\emph {\bibinfo {title}
  {Numerical Recipes 3rd Edition: The Art of Scientific Computing}}}\ (\bibinfo
   {publisher} {Cambridge University Press},\ \bibinfo {address} {New York,
  USA},\ \bibinfo {year} {2007})\BibitemShut {NoStop}%
\bibitem [{\citenamefont {Roland}\ and\ \citenamefont
  {Cerf}(2002)}]{LQA_Roland}%
  \BibitemOpen
  \bibfield  {author} {\bibinfo {author} {\bibfnamefont {J.}~\bibnamefont
  {Roland}}\ and\ \bibinfo {author} {\bibfnamefont {N.~J.}\ \bibnamefont
  {Cerf}},\ }\bibfield  {title} {\emph {\bibinfo {title} {Quantum search by
  local adiabatic evolution},}\ }\href
  {https://doi.org/10.1103/PhysRevA.65.042308} {\bibfield  {journal} {\bibinfo
  {journal} {Phys. Rev. A}\ }\textbf {\bibinfo {volume} {65}},\ \bibinfo
  {pages} {042308} (\bibinfo {year} {2002})}\BibitemShut {NoStop}%
\bibitem [{\citenamefont {Suzuki}(1993)}]{SUZUKI1993}%
  \BibitemOpen
  \bibfield  {author} {\bibinfo {author} {\bibfnamefont {M.}~\bibnamefont
  {Suzuki}},\ }\bibfield  {title} {\emph {\bibinfo {title} {General
  Decomposition Theory of Ordered Exponentials},}\ }\href
  {https://doi.org/10.2183/pjab.69.161} {\bibfield  {journal} {\bibinfo
  {journal} {Proc. Japan Acad. B}\ }\textbf {\bibinfo {volume} {69}},\ \bibinfo
  {pages} {161} (\bibinfo {year} {1993})}\BibitemShut {NoStop}%
\bibitem [{\citenamefont {{De Raedt}}\ and\ \citenamefont
  {Michielsen}(2006)}]{Raedt2006}%
  \BibitemOpen
  \bibfield  {author} {\bibinfo {author} {\bibfnamefont {H.}~\bibnamefont {{De
  Raedt}}}\ and\ \bibinfo {author} {\bibfnamefont {K.}~\bibnamefont
  {Michielsen}},\ }\bibfield  {title} {\emph {\bibinfo {title} {Computational
  Methods for Simulating Quantum Computers},}\ }in\ \href
  {http://arxiv.org/abs/quant-ph/0406210} {\emph {\bibinfo {booktitle} {Quantum
  and Molecular Computing, Quantum Simulations}}},\ \bibinfo {series} {Handbook
  of Theoretical and Computational Nanotechnology}, Vol.~\bibinfo {volume}
  {3},\ \bibinfo {editor} {edited by\ \bibinfo {editor} {\bibfnamefont
  {M.}~\bibnamefont {Rieth}}\ and\ \bibinfo {editor} {\bibfnamefont
  {W.}~\bibnamefont {Schommers}}}\ (\bibinfo  {publisher} {American
  Scientific},\ \bibinfo {address} {Los Angeles},\ \bibinfo {year} {2006})\
  pp.\ \bibinfo {pages} {1--48}\BibitemShut {NoStop}%
\bibitem [{\citenamefont {Young}\ \emph {et~al.}(2008)\citenamefont {Young},
  \citenamefont {Knysh},\ and\ \citenamefont {Smelyanskiy}}]{Young_MEG}%
  \BibitemOpen
  \bibfield  {author} {\bibinfo {author} {\bibfnamefont {A.~P.}\ \bibnamefont
  {Young}}, \bibinfo {author} {\bibfnamefont {S.}~\bibnamefont {Knysh}},\ and\
  \bibinfo {author} {\bibfnamefont {V.~N.}\ \bibnamefont {Smelyanskiy}},\
  }\bibfield  {title} {\emph {\bibinfo {title} {Size Dependence of the Minimum
  Excitation Gap in the Quantum Adiabatic Algorithm},}\ }\href
  {https://doi.org/10.1103/PhysRevLett.101.170503} {\bibfield  {journal}
  {\bibinfo  {journal} {Phys. Rev. Lett.}\ }\textbf {\bibinfo {volume} {101}},\
  \bibinfo {pages} {170503} (\bibinfo {year} {2008})}\BibitemShut {NoStop}%
\bibitem [{\citenamefont {Young}\ \emph {et~al.}(2010)\citenamefont {Young},
  \citenamefont {Knysh},\ and\ \citenamefont {Smelyanskiy}}]{Young_FOPT}%
  \BibitemOpen
  \bibfield  {author} {\bibinfo {author} {\bibfnamefont {A.~P.}\ \bibnamefont
  {Young}}, \bibinfo {author} {\bibfnamefont {S.}~\bibnamefont {Knysh}},\ and\
  \bibinfo {author} {\bibfnamefont {V.~N.}\ \bibnamefont {Smelyanskiy}},\
  }\bibfield  {title} {\emph {\bibinfo {title} {First-Order Phase Transition in
  the Quantum Adiabatic Algorithm},}\ }\href
  {https://doi.org/10.1103/PhysRevLett.104.020502} {\bibfield  {journal}
  {\bibinfo  {journal} {Phys. Rev. Lett.}\ }\textbf {\bibinfo {volume} {104}},\
  \bibinfo {pages} {020502} (\bibinfo {year} {2010})}\BibitemShut {NoStop}%
\bibitem [{\citenamefont {Schulz}(2022)}]{GQW_Master_Thesis}%
  \BibitemOpen
  \bibfield  {author} {\bibinfo {author} {\bibfnamefont {S.}~\bibnamefont
  {Schulz}},\ }\emph {\bibinfo {title} {GPU-accelerated simulation of guided
  quantum walks}},\ \href
  {http://juser.fz-juelich.de/record/916752/files/Master_thesis.pdf} {Master's
  thesis},\ \bibinfo  {school} {RWTH Aachen} (\bibinfo {year}
  {2022})\BibitemShut {NoStop}%
\end{thebibliography}%

\end{document}